\documentclass[12pt,a4paper]{article}
\pdfoutput=1
\usepackage{graphicx}
\usepackage[symbol]{footmisc}

\usepackage{amsmath,amsfonts,amssymb}
\usepackage{cite}
\usepackage{colortbl}
\usepackage{pifont}

\providecommand{\openone}{\leavevmode\hbox{\small1\kern-3.8pt\normalsize1}}

\newcommand{\ptj}{p_{TJ}}

\begin{document}

\vspace*{-2.5cm}
\begin{flushright}
IFT-UAM/CSIC-20-11 \\
\end{flushright}
\vspace{0.cm}

\begin{center}
\begin{Large}
{\bf Jet tagging made easy}
\end{Large}

\vspace{0.5cm}
\renewcommand*{\thefootnote}{\fnsymbol{footnote}}
\setcounter{footnote}{1}
J.~A.~Aguilar-Saavedra$^{a,}$\footnote{On leave of absence from Universidad de Granada, E-18071 Granada, Spain.}, B. Zald\'{\i}var$^{a,b}$ \\[1mm]
\begin{small}
{$^a$ Instituto de F\'isica Te\'orica UAM-CSIC, Campus de Cantoblanco, E-28049 Madrid, Spain} \\ 
{$^b$ Departamento de F\'{\i}sica Te\'orica, Universidad Aut\'onoma de Madrid, Cantoblanco, E-28049 Madrid, Spain}
\end{small}
\end{center}
\renewcommand*{\thefootnote}{\arabic{footnote}}\setcounter{footnote}{0}%

\begin{abstract}
We develop taggers for multi-pronged jets that are simple functions of jet substructure (so-called `subjettiness') variables. These taggers can be approximately decorrelated from the jet mass in a quite simple way. Specifically, we use a Logistic Regression Design (LoRD) which, even being one of the simplest machine learning classifiers, shows a performance which surpasses that of simple variables used by the ATLAS and CMS Collaborations and is not far from more complex models based on neural networks. Contrary to the latter, our method allows for an easy implementation of tagging tasks by providing a simple and interpretable analytical formula with already optimised parameters.      

\end{abstract}

\section{Introduction}

The tagging of boosted jets is an essential tool for the search of heavy new physics beyond the Standard Model (SM) that involves one or more coloured particles as final state. After the first works on jet substructure~\cite{Butterworth:2008iy,Thaler:2008ju,Almeida:2008yp} many jet substructure variables~\cite{Thaler:2010tr,Jankowiak:2011qa,Thaler:2011gf,Larkoski:2013eya,Moult:2016cvt} have been proposed to discriminate boosted Higgs bosons, top quarks and weak $W/Z$ bosons from the QCD background composed of quark and gluon jets. (See also Refs.~\cite{Kaplan:2008ie,Plehn:2009rk,Plehn:2010st}.) Multivariate taggers based on jet substructure variables~\cite{Datta:2017rhs,Moore:2018lsr} or directly using jet images (see Ref.~\cite{Larkoski:2017jix} for a review) have also been employed. These taggers can be qualified as `dedicated', namely they are designed to discriminate jets corresponding to a specific type of signal (Higgs, top, $W/Z$) from the background. Despite achieving a good discrimination (especially for multivariate taggers), a drawback of dedicated taggers is the fact that they may not be sensitive to other types of jets different from those  which they are designed for. This was made clear in Ref.~\cite{Aguilar-Saavedra:2017zuc}: the application of a {\it wrong} tagger to a four-pronged jet can cause the signal significance to plummet, even below the value when the tagger is not applied.

A first attempt to develop a `generic' tagger, sensitive to a variety of complex jets, was done in Ref.~\cite{Aguilar-Saavedra:2017rzt}, by training a neural network (NN) with two, three and four-pronged jets as signals, against the QCD background. Another complementary approach is brought by  weakly-supervised and unsupervised methods, where little or no theoretical input is given and the NN learns to discriminate potential signals in different ways, for example by comparing different kinematical regions~\cite{Collins:2019jip,Nachman:2020lpy,Andreassen:2020nkr} or by a reduction of the space dimensionality using an autoencoder~\cite{Farina:2018fyg,Heimel:2018mkt,Roy:2019jae,Cerri:2018anq}. Despite all these multivariate methods are very effective in the discrimination, they have two inherent drawbacks:
\begin{itemize}
\item Often, there is no obvious interpretation of how the multivariate methods work in separating signal from background. This is because the interpretability of the results greatly decreases in general with the complexity of the machine learning (ML) models.

\item More complex ML models require more complex implementations, which translates into more difficult testing and reproducibility of the results, especially when the literature does not provide all the necessary details of the implementation (which is typically the case, unfortunately).
\end{itemize}
As a complement to these methods, it is our purpose here to develop and test a set of simple taggers for multi-pronged jets, based on a set of variables proposed in Ref.~\cite{Datta:2017rhs},
 \begin{equation}
 \left\{ \tau_1^{(1/2)}, \tau_1^{(1)}, \tau_1^{(2)}, \dots , \tau_{M-2}^{(1/2)}, \tau_{M-2}^{(1)}, \tau_{M-2}^{(2)}, \tau_{M-1}^{(1)}, \tau_{M-1}^{(2)} \right\} \,,
 \label{ec:taulist}
 \end{equation}
with $M > 1$ an integer. 
 The subjettiness variables $\tau_n^{(\beta)}$ measure to which extent the radiation in a jet is clustered around $n$ axes, with $\beta$ an angular exponent (their precise definition is given in the next section).  As shown in Ref.~\cite{Datta:2017rhs}, the set of variables (\ref{ec:taulist}) allows to reconstruct the phase space of $M$ partons within a jet, up to a global rotation. 

Our main goal in this work is simplicity: the taggers obtained are simple functions of the subjettiness variables (\ref{ec:taulist}), with an approximate mass decorrelation \`a la DDT~\cite{Dolen:2016kst} which suffices to maintain the shape of the jet mass distribution to a large extent, after the application of the taggers. The taggers are developed using Logistic Regression Design (LoRD) to find the numerical parameters in the function that achieve the best discrimination, as described in Sect.~\ref{sec:2} and appendix~\ref{app:LR}. There exist already in the literature taggers based on optimised products of subjettiness variables:  a method to develop taggers with a scan over the parameter space was earlier introduced in Ref.~\cite{Datta:2017lxt}, and complex neural networks were used in Ref.~\cite{Datta:2019ndh}, for the development of taggers to discriminate two-pronged jets versus QCD jets, and for quark/gluon jet identification. With respect to those, the design of our taggers is much simpler, and also addresses the two issues mentioned above about interpretability and reproducibility.

We find that the discrimination power of our simple taggers in some cases largely surpasses the simple variables used by the ATLAS and CMS Collaborations in searches for new physics using jet substructure. Results are presented in Sect.~\ref{sec:3}. We also compare the results using LoRD with NNs using the same architecture as in Ref.~\cite{Aguilar-Saavedra:2017rzt}. In general, the NNs perform better, except for some signals which neither are trained for.

An important point in the design of the taggers is the kinematical region (i.e. jet mass and $p_T$) used for the optimisation. We address this issue in Sect.~\ref{sec:4}. While the dependence on $\ptj$ is marginal, the dependence on jet mass is more noticeable when one gets away from the design region. Our conclusions are presented in Sect.~\ref{sec:5}. In appendix~\ref{sec:var} we estimate the variance of the taggers obtained with the LoRD. A qualitative discussion about the interpretability and the intrinsic dimension of the datasets is discussed in appendix \ref{sec:PCA}. In appendix~\ref{sec:newa} we compare among different options for the design of the taggers, regarding the grooming (or not) of the jet mass, momentum and subjettiness variables. In appendix~\ref{sec:a} we summarise a few results for taggers without mass decorrelation, and in appendix~\ref{sec:b} we investigate the performance of the more complex taggers --- the ones designed for four-pronged jets --- for jets with less prongs.

\section{LoRD of the taggers}
\label{sec:2}

The input to the taggers is given by a set of subjettiness variables (\ref{ec:taulist}) with $M\leq 9$,\footnote{In Ref.~\cite{Aguilar-Saavedra:2017rzt} it was shown that for the topologies considered in this work (2-, 3- and 4-pronged) going beyond $M=7$ does not contribute further to the discrimination performance of a NN. With Logistic Regression we find that the discrimination power saturates at $M=9$.} where 
 \begin{equation}
 \tau_n^{(\beta)} = \frac{1}{\ptj} \sum_{i} p_{Ti} \; \text{min} \left\{ \Delta R_{1i}^\beta, \Delta R_{2i}^\beta, \dots, \Delta R_{ni}^\beta \right\} \,,
 \label{ec:tauN}
 \end{equation}
  with $i$ labelling the particles in the jet, $p_{Ti}$ their transverse momenta, $\Delta R_{Ki} $ their lego-plot distance to the axis $K=1,\dots,N$ and $\ptj$ the jet transverse momentum. As in Ref.~\cite{Datta:2017rhs}, in the computation of these variables we use the axes defined by exclusive $k_T$ algorithm~\cite{Catani:1993hr,Ellis:1993tq} with standard $E$-scheme recombination~\cite{Blazey:2000qt}. 

The proposed functional form for the taggers is
\begin{equation}
T = \bar T - b \rho -a \,,
\label{ec:T}
\end{equation}
with 
\begin{equation}
\bar T = \sum_{n,\beta} c_n^\beta \log \tau_n^{(\beta)}
\label{ec:Tbar}
\end{equation}
and $\rho=\log m_{J}^2 / \ptj^2$. The coefficients $c_n^\beta$ are determined by Logistic Regression to optimise the discrimination between the signal(s) and the background (see appendix \ref{app:LR} for a detailed description of this implementation); $b$ is a parameter chosen to achieve an approximate mass decorrelation, and for convenience the tagger output is shifted by subtracting a fixed quantity $a$ so that its average $\langle T \rangle$, when evaluated on a reference background sample, vanishes. Notice that the sum (\ref{ec:Tbar}) is equivalent to a product
\begin{equation}
 \prod_{n,\beta} \left( \tau_n^{(\beta)} \right) ^{c_n^\beta} \,,
 \label{ec_Tbarprod}
 \end{equation}
 that includes and generalises the commonly used ratios $\tau_{21} = \tau_2^{(1)} / \tau_1^{(1)}$ and $\tau_{32} = \tau_3^{(1)} / \tau_2^{(1)}$.
Several comments and clarifications regarding our procedure are now in order.
\begin{itemize}
\item Independently of the precise method used to determine $c_n^\beta$, a range of jet mass and $p_T$ has to be specified for the signal(s) and background. For any of these processes the distributions of the variables $\tau_n^{(\beta)}$ depend on $m_J$ and $\ptj$. The coefficients $c_n^\beta$ are then obtained to optimise the discrimination of signal and background within a given interval of jet mass and $p_T$.
\item Using ungroomed jet mass and $p_T$ for the determination of $c_n^\beta$ reduces the dependence of the resulting taggers on the intervals chosen, and slightly improves the mass decorrelation. Also, this is desirable in order not to stick to a particular grooming algorithm. We have also tried using the groomed mass and $p_T$, and the discrimination power is similar. We also use $\tau_n^{(\beta)}$ of the ungroomed jets, since we find that the discrimination between signal and background is better. A comparison among these possibilities is made in appendix~\ref{sec:newa}
\item For the mass decorrelation, evaluation of the tagger performance, etc., namely in all calculations except the tagger design itself, we use $m_J$ and $\ptj$ of the groomed jets. The recursive soft drop~\cite{Dreyer:2018tjj} algorithm with parameters $\beta = 1$, $z_\text{cut} = 0.05$, $N=3$ is found to work very well for multi-pronged jets, avoiding the peak distortions and shifts that other algorithms produce~\cite{Aguilar-Saavedra:2018xpl}.
\item The parameter $a$ is chosen to adjust $\langle T \rangle = 0$ for a reference background sample with (groomed) $\ptj \geq 250$ GeV, and without any restriction on $m_J$. For larger $p_T$ or when considering a narrow $m_J$ interval, the average is slightly shifted. This residual dependence may be accounted for by varying the tagger threshold, as done for example in Ref.~\cite{Sirunyan:2017nvi} or, equivalently, by varying $a$ as a function of $m_J$ and $\ptj$. This sophistication however is not required for our discussion.
\end{itemize}

With the LoRD method we optimise the discrimination between quark/gluon jets (background) and multi-pronged decays of boosted colour singlet particles (signal). Quark and gluon jets are obtained by generating the parton-level processes $p p \to Zg$, $p p \to Zq$, with decay  $Z \to \nu \nu$, using {\scshape MadGraph5}~\cite{Alwall:2014hca}, and {\scshape Pythia~8}~\cite{Sjostrand:2007gs}. In all cases the centre-of-mass energy is set to 13 TeV. The detector response is simulated with {\scshape Delphes 3.4}~\cite{deFavereau:2013fsa} using the CMS detector card. Jets are reconstructed using the anti-$k_T$ algorithm~\cite{Cacciari:2008gp}  with radius $R=0.8$. {\scshape FastJet 3.2}~\cite{Cacciari:2011ma} is used for jet reconstruction, grooming and calculation of the $\tau_n^{(\beta)}$ variables. For the signal we use fat jets resulting from the decay of neutral, colour-singlet particles into four, three and two quarks, considering the six processes
\begin{align}
& p p \to Z' \to S \, Z(\to \nu \nu) \,, && S \to u \bar u u \bar u ~~ \text{ and } ~~ S \to b \bar{b} b \bar{b}, \notag \\
& p p \to Z' \to F \,\nu\, Z ( \to \nu \nu)  \,, && F \to u d d ~~ \text{ and } ~~ F \to u b b, \notag \\
& p p \to Z' \to S \, Z(\to \nu \nu) \,, && S \to u \bar u ~~ \text{ and } ~~ S \to b \bar b ,
\label{ec:MIdata}
\end{align}
with $S$ a scalar and $F$ a fermion. These processes are generated at parton level with {\scshape Protos}~\cite{protos} and, subsequently passed through the parton shower, hadronisation and fast simulation chain.
In order to remain as model-agnostic as possible, we implement decays of $S$ and $F$ with a flat matrix element, so that the decay weight of the different kinematical configurations only corresponds to the two-, three- or four-body phase space. These signal Monte Carlo data are dubbed as Model Independent (MI) data, and its use is motivated by the need to sample phase space without model prejudice~\cite{Aguilar-Saavedra:2017rzt}. This choice is very effective to make the taggers learn prongness rather than other undesired feature. Consequently, the obtained taggers $T$ can be used outside --- though not very far from --- the interval they have been designed for.

We build taggers for four-pronged (4P), three-pronged (3P) and two-pronged (2P) signals using the corresponding set of signal processes in the first, second and third line of Eqs.~(\ref{ec:MIdata}) as signal, versus the QCD background.
For 4P taggers we select $M=9$ in (\ref{ec:taulist}), while for 3P and 2P taggers it is enough to use $M=7$ and $M=5$, respectively.\footnote{We have also examined the possibility of generic taggers using all signal sets at once. Their performance is worse, especially on four-pronged signals of high mass.} We select three different kinematical regions (for ungroomed quantities) for the design of the taggers, labelled as follows:
\begin{itemize}
\item {\tt hi80}: $\ptj \geq 1$ TeV, $m_J \in [60,100]$ GeV.
\item {\tt hi200}: $\ptj \geq 1$ TeV, $m_J \in [170,230]$ GeV.
\item {\tt lo80}: $\ptj \geq 500$ GeV, $m_J \in [60,100]$ GeV.
\end{itemize}
For the backgrounds we set at the parton level a cut on the jet, $p_T \geq 1$ TeV (for {\tt hi80} and {\tt hi200}) and $p_T \geq 500$ GeV (for {\tt lo80}) in order to increase the efficiency of the event generation. After simulation, the cuts on $\ptj$ and $m_J$ are applied. 
For the signals we set $M_{Z'} = 2.2$ TeV for {\tt hi80} and {\tt hi200}, and $M_{Z'} = 1.1$ TeV for {\tt lo80}, in order to have a transverse momentum distribution close to the one that is subsequently required by the cut on $\ptj$. The mass of the intermediate particles $S$, $F$ is set to 80 GeV for {\tt hi80} and {\tt lo80} and 200 GeV for {\tt hi200}. After the simulation, the cut on $m_J$ is applied to select the corresponding kinematical region.

In total, we develop nine taggers (2P, 3P and 4P for the three kinematical regions defined above), plus three alternate versions of the {\tt hi80} taggers for testing purposes. The size of the signal and background datasets used in the optimisation is collected in Table~\ref{tab:train}. The background events are approximately evenly divided among quark and gluons. The signal events for each (4P, 3P, 2P) signal class are approximately evenly divided among the two contributions listed in each line of Eqs.~(\ref{ec:MIdata}). Finally, the two classes (signal and background) are mutually balanced among each other as well. The results for the coefficients for the different taggers are collected in Table~\ref{tab:coef}. The variance of the obtained results is addressed in appendix~\ref{sec:var}, while a qualitative discussion about the interpretability and the intrinsic dimension of the datasets is discussed in appendix \ref{sec:PCA}. 

\begin{table}[htb]
\begin{center}
\begin{tabular}{lccccc}
& background & 4P signal & 3P signal & 2P signal \\
{\tt hi80}   & 147869 & 55695 & 45830 & 54004 \\
{\tt hi200} & 233889 & 66717 & 63257 & 72282 \\
{\tt lo80}   & 323175 & 65011 & 52431 & 58279 
\end{tabular}
\end{center}
\caption{Number of events used in the optimisation and test of the taggers with the LoRD method.}
\label{tab:train}
\end{table}

For the mass decorrelation and test of the tagger performance we use a sample of QCD dijet production generated with {\scshape MadGraph} in 100 GeV intervals of $p_T$, starting at $[200,300]$ GeV and with the last one having $p_T \geq 2.2$ TeV. The different samples are hadronised and passed through the detector simulation, and then combined with a weight that corresponds to the cross section. For each interval we generate $2 \times 10^5$ events and keep both jets (leading and sub-leading) for the analysis, therefore our QCD sample comprises $8.4$ million jets, spanning a very wide range of mass and $p_T$.

\begin{table}[htb]
\begin{center}
\begin{tabular}{cccccccccc}
& \multicolumn{3}{c}{\tt hi80} & \multicolumn{3}{c}{\tt hi200}  & \multicolumn{3}{c}{\tt lo80} \\
& $T_\text{4P}$ & $T_\text{3P}$ & $T_\text{2P}$ 
& $T_\text{4P}$ & $T_\text{3P}$ & $T_\text{2P}$
& $T_\text{4P}$ & $T_\text{3P}$ & $T_\text{2P}$ \\
$c_1^1$ & 1.839 & 2.792 & 2.773 & 2.133 & 2.759 & 3.000 & 2.334 & 2.665 & 2.863
\\
$c_2^1$ & 1.092 & 1.411 & -0.240 & 1.876 & 1.392 & 0.083 & 1.629 & 1.076 & 0.020
\\
$c_3^1$ & 0.448 & 0.158 & 0.436 & 0.841 & 0.345 & -0.091 & 0.630 & 0.209 & 0.248
\\
$c_4^1$ & 0.023 & 0.048 & -0.797 & 0.106 & -0.419 & -0.706 & 0.311 & -0.075 & -0.083
\\
$c_5^1$ & 0.608 & 0.369 & 0         & 0.070 & -0.675 & 0        & -0.156 & -0.079 & 0
\\
$c_6^1$ & 0.274 & -0.094 &  0       & -0.017 & -0.499 & 0        & 0.134 & 0.344 & 0
\\
$c_7^1$ & 0.221 & 0        & 0         & -0.208 & 0        & 0        & 0.397 & 0         & 0
\\
$c_8^1$ & 0.477 & 0        & 0         & 0.048 & 0        & 0         & -0.380 & 0         & 0 
\\
$c_1^2$ & 0.168  & 1.097  & 1.941  & 0.533 & 0.734  & 2.470 & 0.850 & 1.412 & 1.874
\\
$c_2^2$ & -0.458 & -0.854 & -1.151 & 1.049 & 0.607 & -1.071 & 0.059 & -0.436 & -1.746
\\
$c_3^2$ & -1.132 & -0.726 & -1.231 & -0.197 & -0.407 & -1.095 & -0.428 & -0.707 & -0.668
\\
$c_4^2$ & -1.058 & -1.468 & -1.494 & -0.588 & -0.983 & -1.310 & -1.108 & -1.266 & -1.134
\\
$c_5^2$ & -0.700 & -1.093 & 0         & -0.865 & -1.045 & 0         & -0.917 & -0.786 & 0
\\
$c_6^2$ & -0.864 & -0.413 & 0         & -0.943 & -0.821 & 0         & -1.250 & -1.025 & 0
\\
$c_7^2$ & -0.308 & 0        & 0         & -0.710 & 0        & 0          & -0.256 & 0          & 0  
\\
$c_8^2$ & -0.424 & 0        & 0         & -1.203 &  0        & 0         & -1.047 & 0          & 0  
\\
$c_1^{0.5}$ & 2.157 & 2.343 & 3.262 & 2.298 & 3.020 & 2.539 & 2.514 & 2.814 & 3.113
\\
$c_2^{0.5}$ & 1.615 & 1.478 & 0.302 & 2.011 & 1.777 & -0.135 & 1.893 & 1.688 & -0.101
\\
$c_3^{0.5}$ & 1.095 & 1.071 & -0.112 & 0.711 & 0.359 & 0.300 & 0.953 & 0.288 & -0.255
\\
$c_4^{0.5}$ & 1.002 & -0.021 & 0       & 0.452 & 0.370 & 0        & 0.494 & 0.778 & 0
\\
$c_5^{0.5}$ & 0.561 & 0.255 & 0        & 0.381 & 0.155 & 0        & 0.747 & 0.158 & 0
\\
$c_6^{0.5}$ & 0.403 & 0        & 0        & -0.002 & 0        & 0        & 0.959 & 0        & 0
\\
$c_7^{0.5}$ & 0.358 & 0        & 0        & -0.083 & 0        & 0        & 0.853 & 0        & 0 
\\
$b$ & 0.731 & 1.008 & 0.978 & 0.984 & 1.059 & 1.161 & 1.087 & 1.148 & 1.044
\\
$a$ & 3.093 & 4.243 & 5.382 & 3.334 & 3.688 & 4.801 & 3.890 & 4.082 & 4.240
\end{tabular}
\end{center}
\caption{Numerical coefficients in (\ref{ec:T}) and (\ref{ec:Tbar}) corresponding to the different taggers.}
\label{tab:coef}
\end{table}

The application of a tight cut on the value of $\bar T$ produces a noticeable modification in the lineshape of the QCD background versus the jet mass, as seen in appendix~\ref{sec:a}. This is a serious inconvenient in experimental searches for a bump in this distribution. For other searches that do not use the jet mass as final discriminator, maintaining the shape is still desirable in order to be able to use sidebands for the estimation of the background. We therefore perform an approximate mass decorrelation following the DDT~\cite{Dolen:2016kst} proposal. For each tagger we select the parameter $b$ by fitting the calculating the average slope of $\langle \bar T \rangle$ versus $\rho$, in the interval $\rho\in[-6,-2]$, for the QCD background sample. Because this average also depends on $\ptj$, we select $\ptj \in[500,600]$ GeV, which gives good results when the dependence on $\rho$ is not linear and the averages show some spread with $\ptj$. Finally, the parameter $a$ is adjusted so as to have $\langle T \rangle = 0$ in the inclusive QCD sample with $\ptj \geq 250$ GeV. The values of $b$ and $a$ for each tagger are collected in the last two lines of Table~\ref{tab:coef}.

\begin{figure}[t]
\begin{center}
\begin{tabular}{ccc}
\includegraphics[height=5.2cm,clip=]{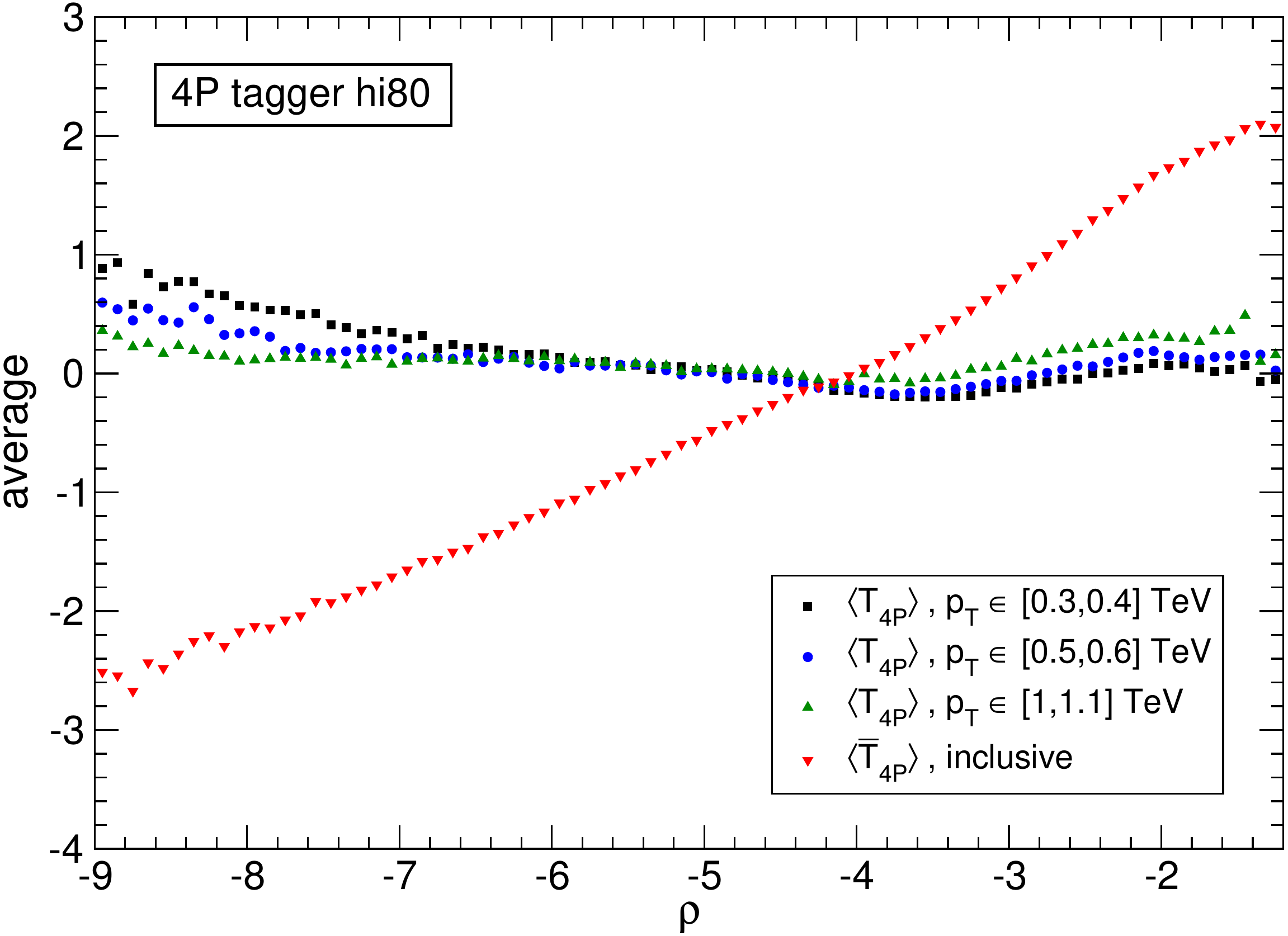} & \quad &
\includegraphics[height=5.2cm,clip=]{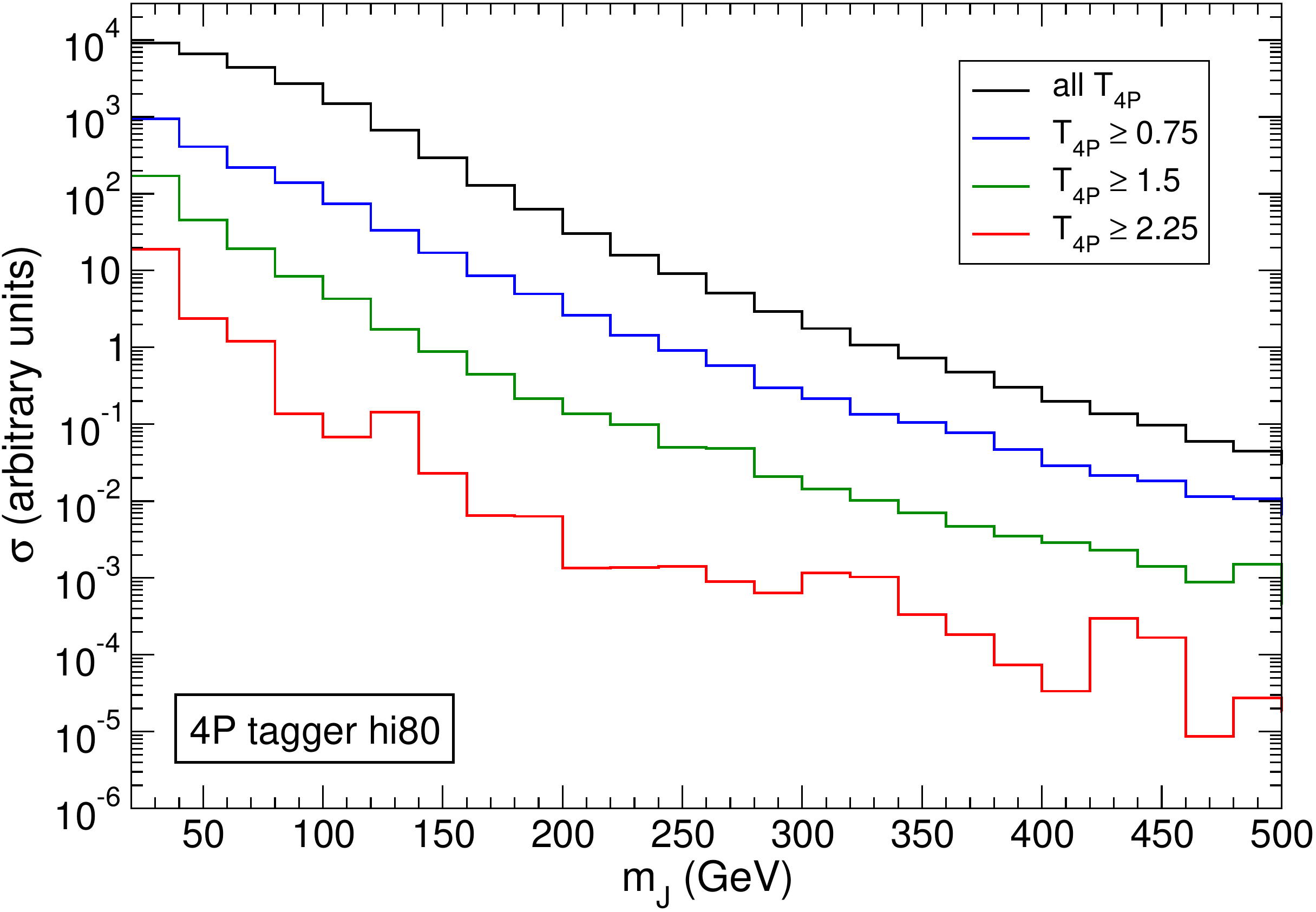} \\
\includegraphics[height=5.2cm,clip=]{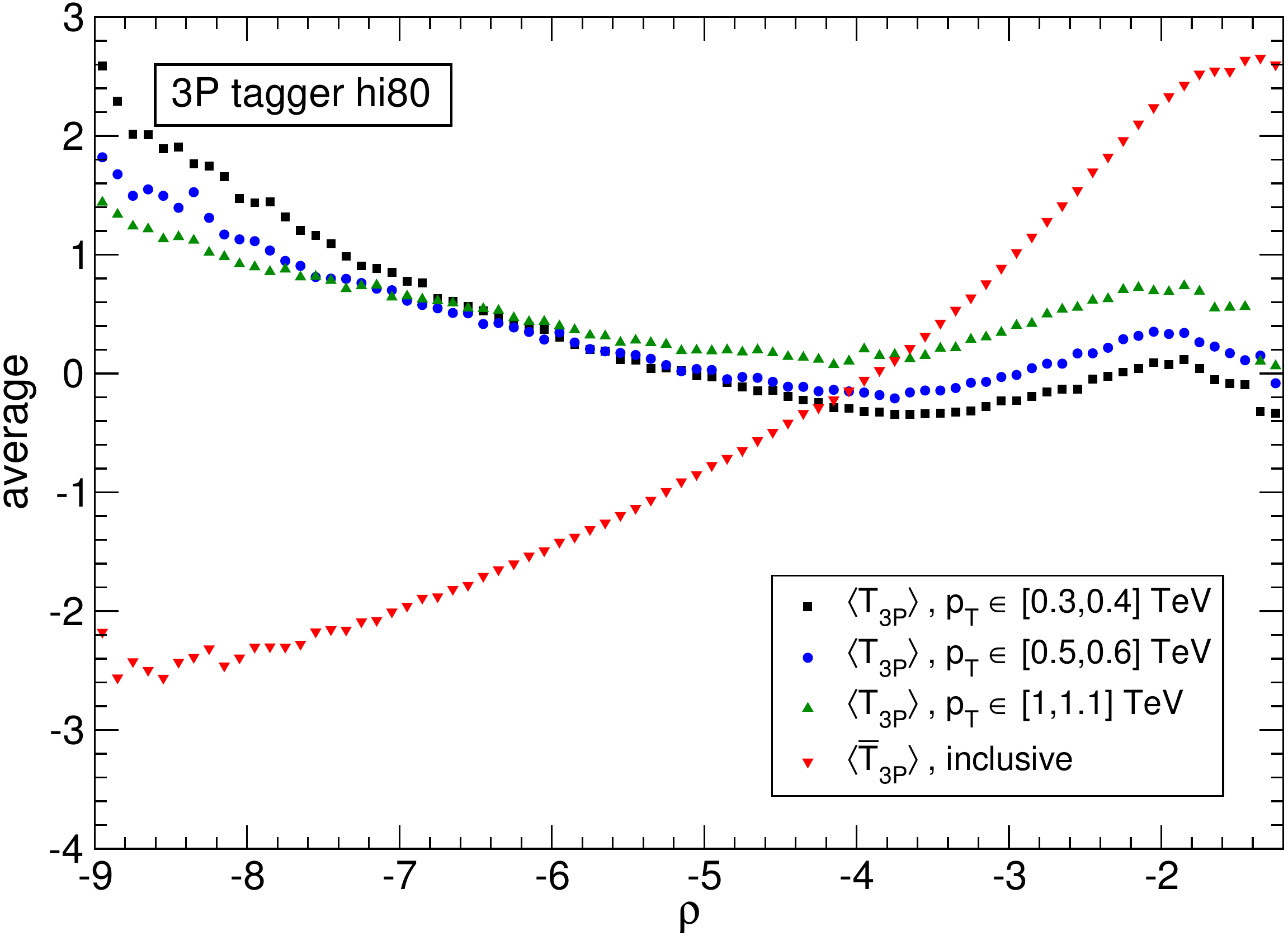} & \quad &
\includegraphics[height=5.2cm,clip=]{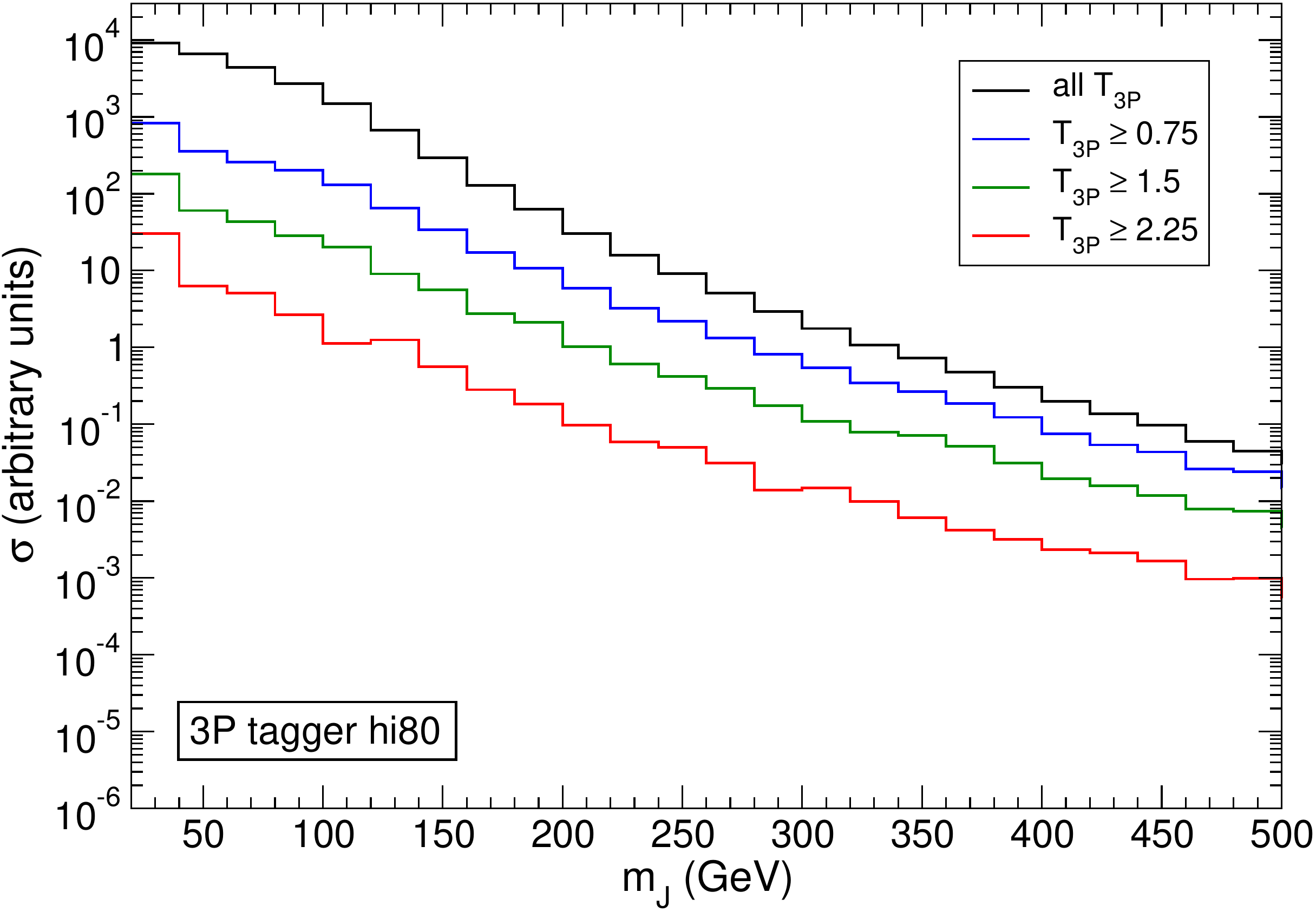} \\
\includegraphics[height=5.2cm,clip=]{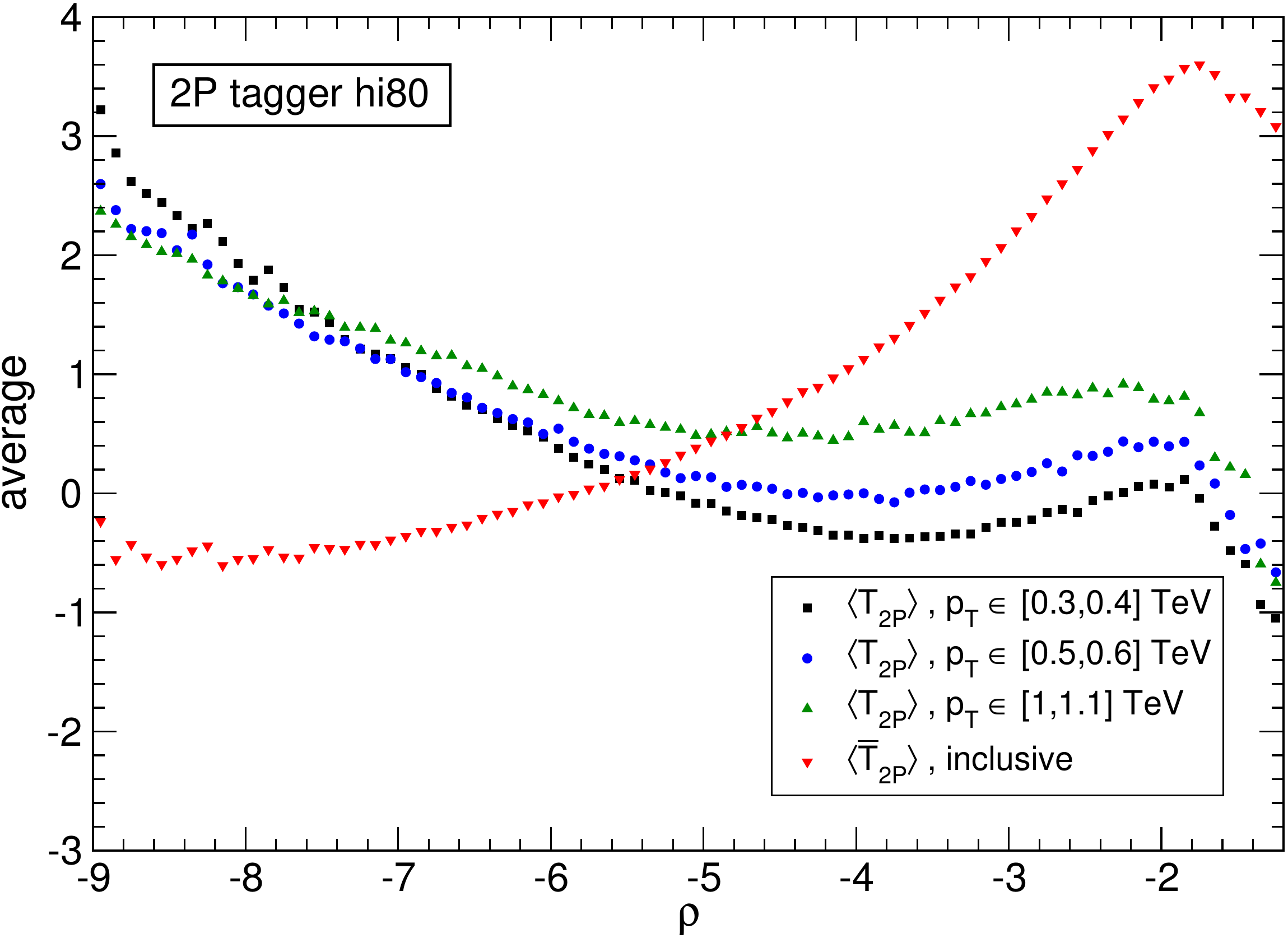} & &
\includegraphics[height=5.2cm,clip=]{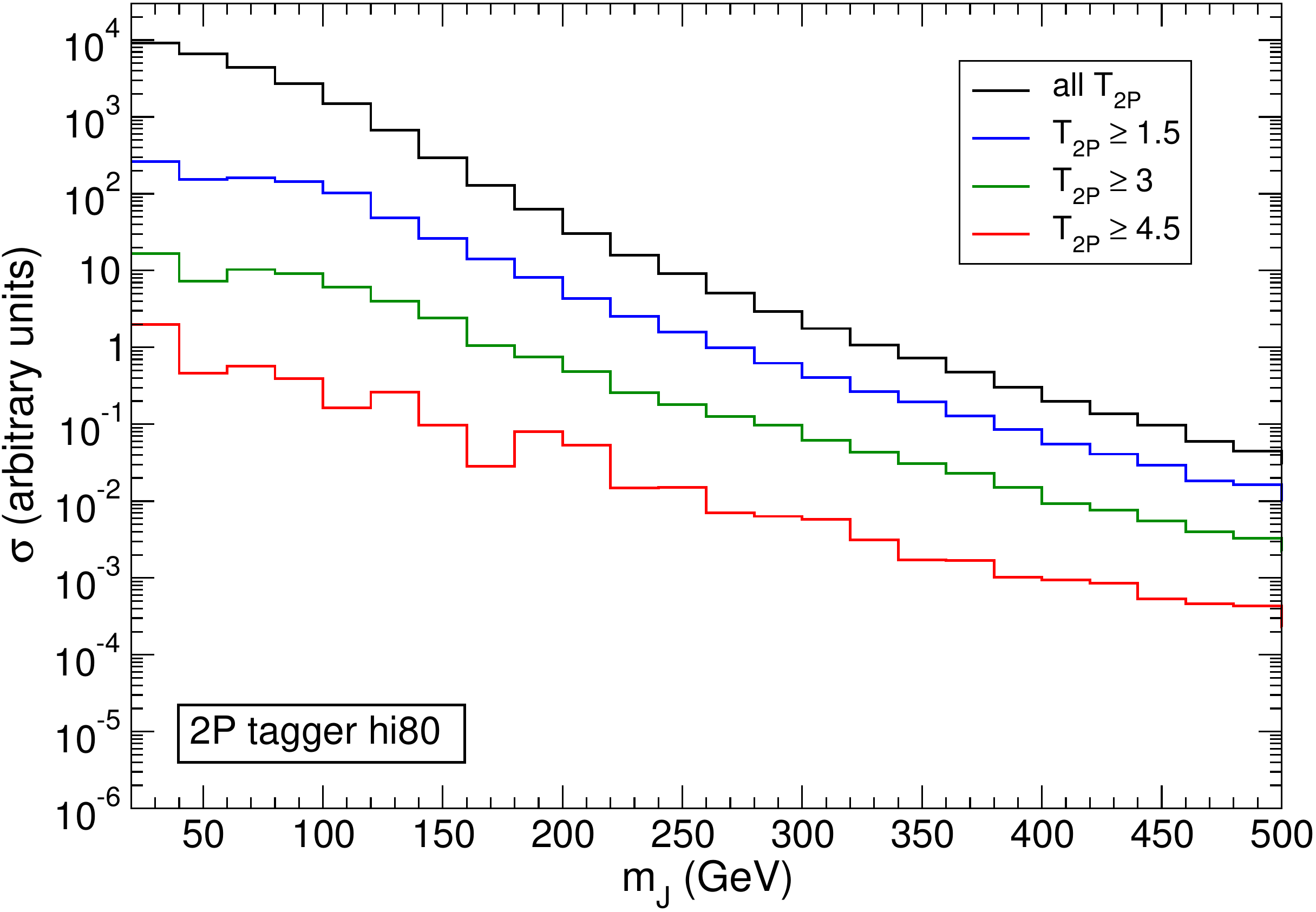} 
\end{tabular}
\caption{Left: average $\langle T \rangle$  of the three mass-decorrelated {\tt hi80} taggers evaluated for the QCD background in three $\ptj$ bins, as a function of $\rho$. For comparison, the average $\langle \bar T \rangle$ in the inclusive QCD sample is shown in red. Right: Jet mass distribution for the QCD background, after increasingly tighter cuts on the tagger output. }
\label{fig:mdecorr}
\end{center}
\end{figure}

For illustration, we show in the left panels of Fig.~\ref{fig:mdecorr} the average $\langle T \rangle$ for the QCD background in several $\ptj$ bins, as a function of $\rho$. The top, middle and bottom panels correspond to the 4P, 3P and 2P {\tt hi80} taggers, respectively. For comparison, we also show $\langle \bar T \rangle$ for the non-decorrelated tagger, for the inclusive sample. For the 4P tagger the mass decorrelation is very good: the three lines for $\langle T \rangle$ are almost coincident and horizontal. For the 3P and 2P taggers the decorrelation achieved with this simple prescription is poorer. The same trend is found with the {\tt hi200}, {\tt lo80} and the alternate {\tt hi80} taggers.
The jet mass distributions for the QCD background after increasingly tighter cuts on $T$ are presented on the right panels of Fig.~\ref{fig:mdecorr}. We observe that the background lineshape is very well preserved, with results comparable to the best decorrelation methods examined Ref.~\cite{Bradshaw:2019ipy}. One can notice two minor features:
\begin{itemize}
\item[(i)] When the background is suppressed by a factor around 100 (e.g. green curve with respect to black curve in Fig.~\ref{fig:mdecorr}-top right), a small increase is produced in the first bin $[20,40]$ GeV.
\item[(ii)] For the 3P and 2P taggers the slope of the QCD distribution (see right plots) slightly decreases after application of the cuts on $T$ (coloured lines). This corresponds to the fact that the $\langle T \rangle$ distributions on the left panels are not as flat as for the 4P tagger.

\end{itemize}
We remark that this simple decorrelation with fixed $b$ in (\ref{ec:T}) can easily be improved by taking $b$ as a function of $m_J$ and $\ptj$. As the numerical calculation of $\bar T$ is very simple, this is a rather computer-inexpensive task. Our purpose here is to show that even a rough mass decorrelation with fixed $b$ does most of the job to maintain the profile of the QCD jet mass distribution after the application of the tagger, even for very tight cuts on the tagger output. Refinements are always possible, see Ref.~\cite{Dolen:2016kst}.

\begin{figure}[t]
\begin{center}
\begin{tabular}{ccc}
\includegraphics[height=4.7cm,clip=]{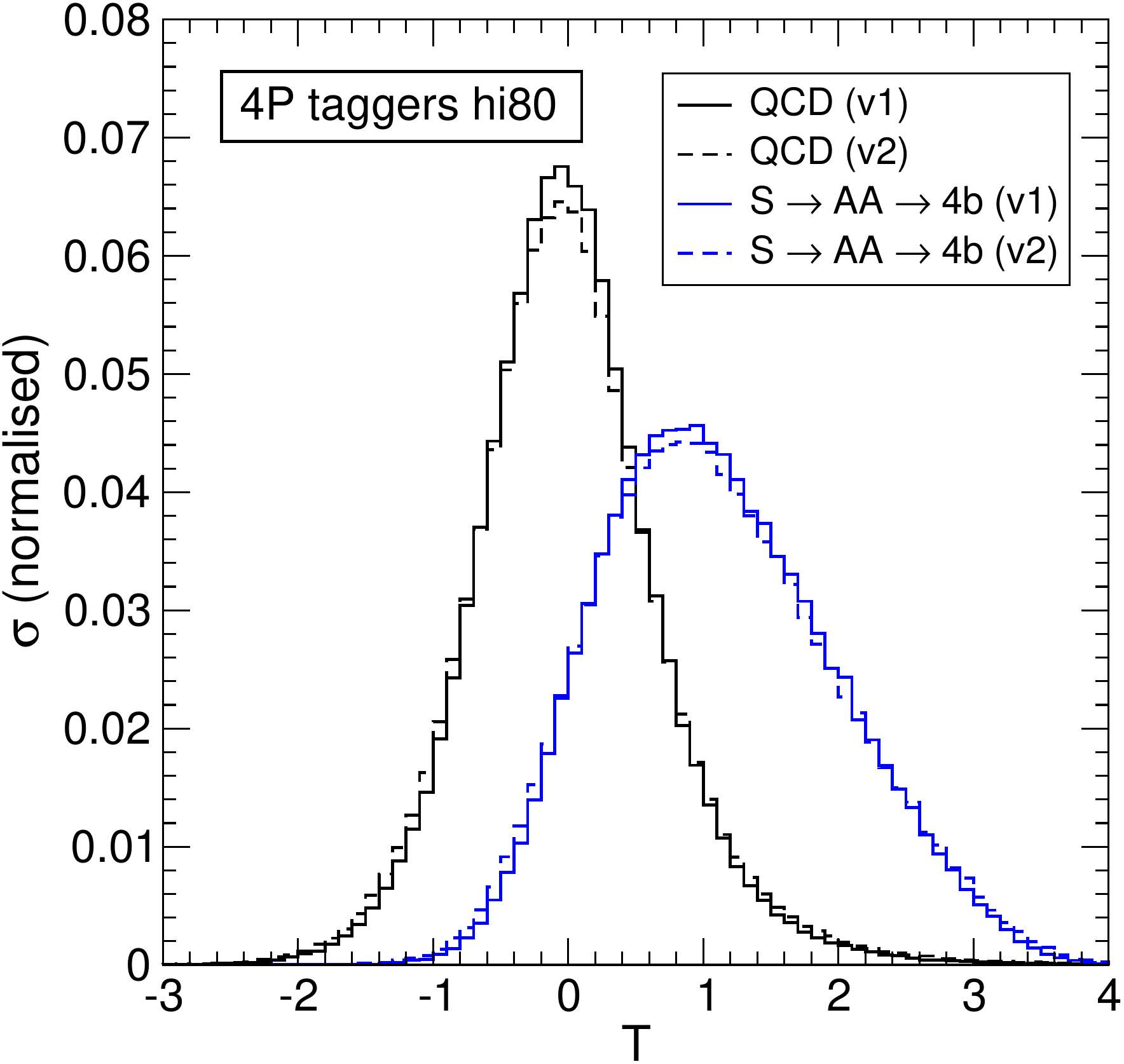} &
\includegraphics[height=4.7cm,clip=]{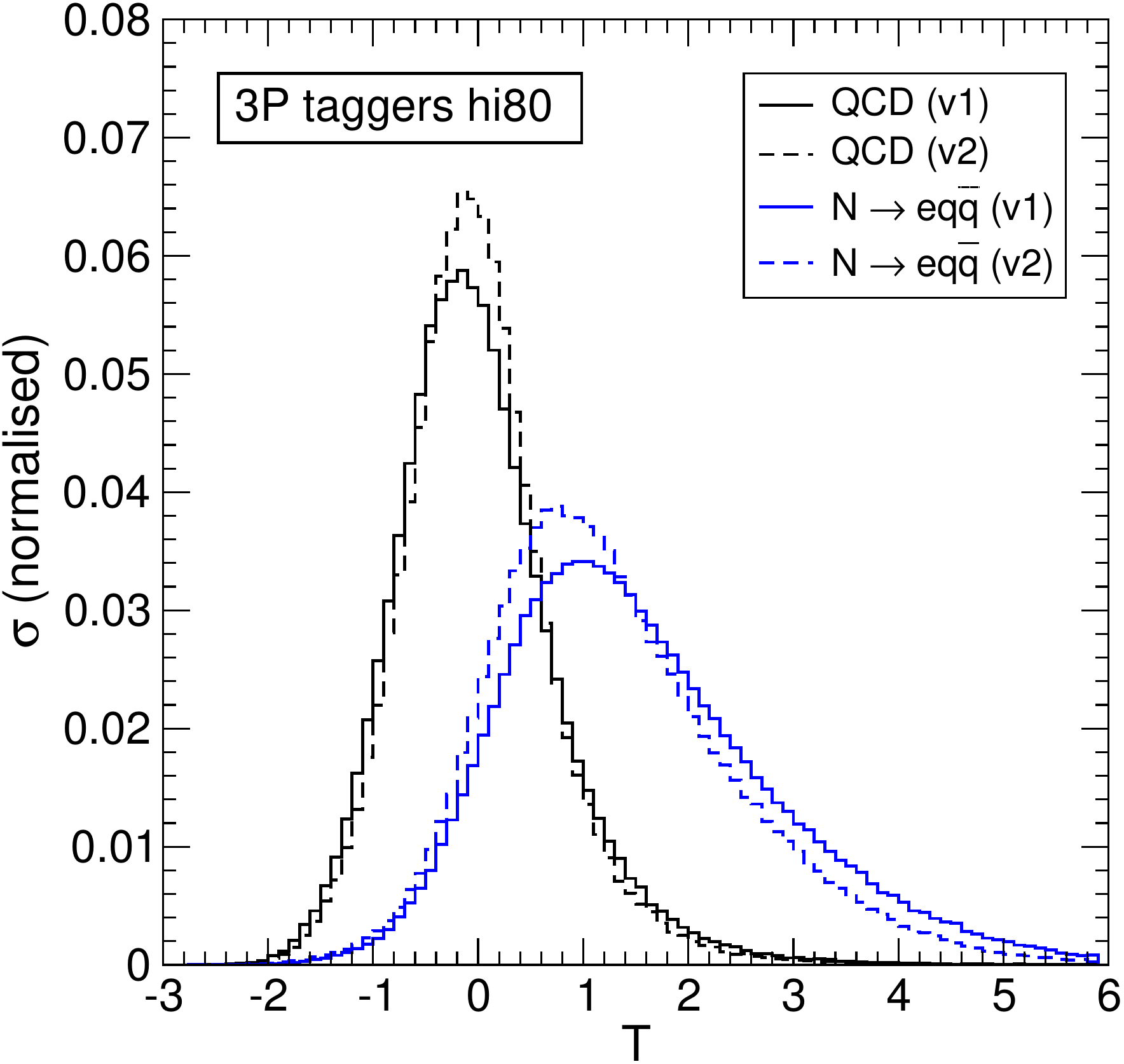} & 
\includegraphics[height=4.7cm,clip=]{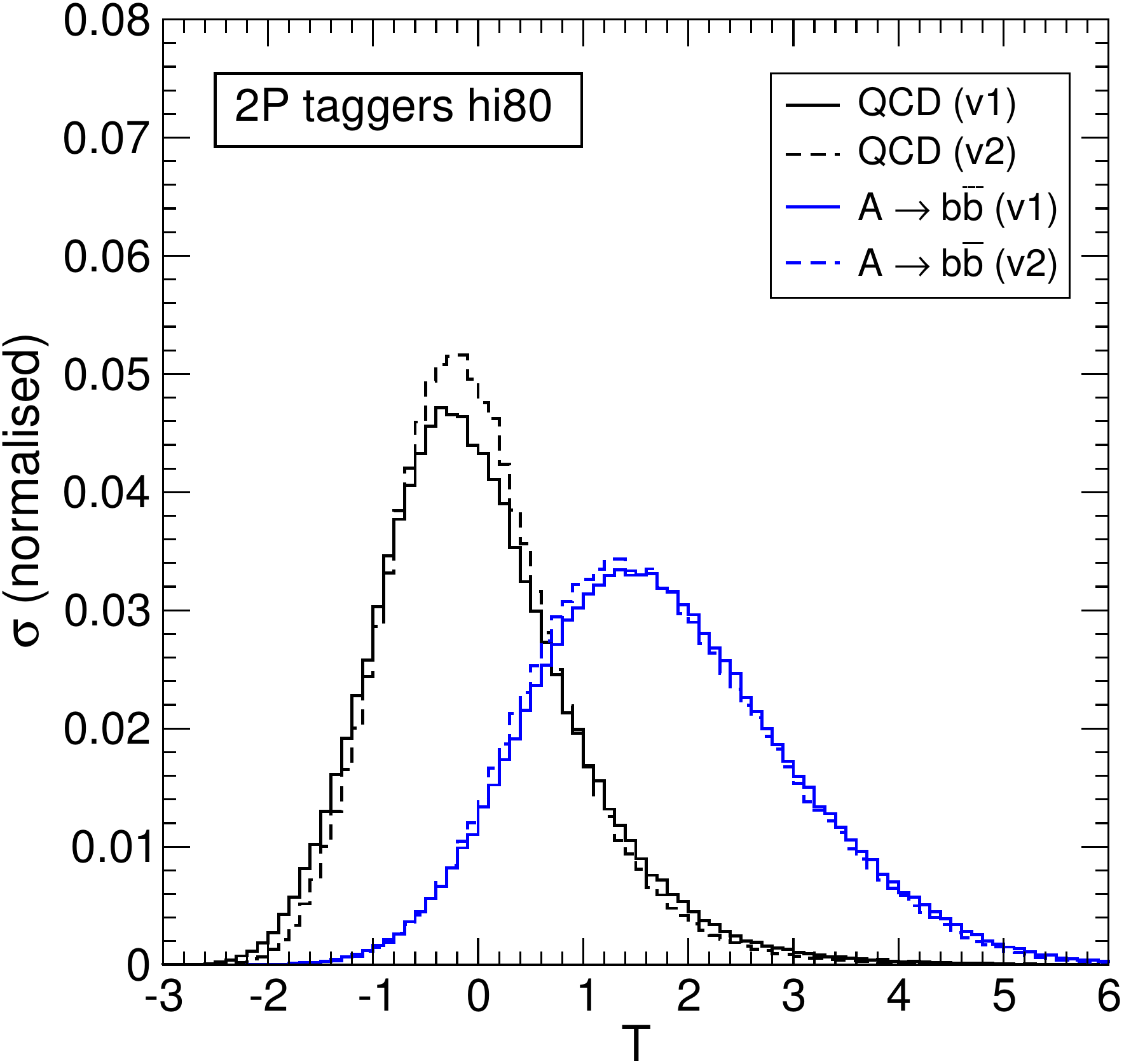} 
\end{tabular}
\caption{Distributions of $T$ for the QCD background (black) and selected signals (blue), for the {\tt hi80} taggers as in Table~\ref{tab:coef} (v1, solid lines) and alternate versions (v2, dashed lines) obtained with a different random seed.}
\label{fig:v1v2}
\end{center}
\end{figure}

A final point that is of interest for practical applications is the {\em stability} of the results obtained by the LoRD. The stability of the classifier performance on test samples is very good, as shown in appendix~\ref{sec:var}. However, this does not guarantee that the $T$ distributions for the background and signals are similar.  With this purpose, we have used an alternate set of {\tt hi80} taggers obtained with different random seeds to check to which extent the $T$ distributions are alike. 
The results, presented in Fig.~\ref{fig:v1v2}, show a remarkable similarity between the alternate versions (v1 and v2) of the same tagger. No cut on jet mass is applied, and $\ptj \geq 250$ GeV is required on signal and background for consistency with the parton-level cut in the background generation. The details on the signals can be found in the next section. This stability is useful to be able to build a set of taggers that cover a very wide range of jet masses, if necessary. Notice also that, by design, the signals are expected to have higher values of $\bar T$ than the background, therefore the background can be reduced with a lower cut on $\bar T$.

\section{Performance}
\label{sec:3}

We evaluate the performance of our taggers by selecting several new physics signals yielding multi-pronged jets,
\begin{align}
& p p \to Z' \to S S \,,\quad S \to AA \to b \bar b b \bar b\,, \notag \\
& p p \to Z' \to S S \,,\quad S \to WW \to q \bar q q \bar q \,, \notag \\
& p p \to Z' \to A A \,,\quad A \to b \bar b \,, \notag \\
& p p \to Z' \to W W \,,\quad W \to q \bar q \,,
\end{align}
with $q=u,d,s,c$ light quarks other than $b$. We generically denote the scalars with cascade decay into four quarks as $S$, and the scalars decaying into $b \bar b$ as $A$.  The decays $Z' \to SS$, $Z' \to AA$ can take place in any SM extension with an additional $\text{U}(1)'$ group and extra scalars, with the minimal consistent implementation given in Ref.~\cite{Aguilar-Saavedra:2019adu}. The decays $Z' \to WW$ can take place in left-right models~\cite{Mohapatra:1974gc,Senjanovic:1975rk}. 
We select $M_{Z'} = 2.2$ TeV and different values of $M_S$ and $M_A$ to test the {\tt hi80} and {\tt hi200} taggers. For three-pronged jets from $Z' \to t \bar t$ we have not found significant improvement over the simple ratio $\tau_{32}$ and we omit the results for brevity. 
The generated signal samples have a minimum of $10^5$ events, and for each event we use both jets, therefore the samples used have a minimum of $2 \times 10^5$ jets.  The background sample is the same one with $8.4 \times 10^6$ events used for the mass decorrelation.

A meaningful assessment of the performance of the taggers can only be made within a given interval of jet mass and $p_T$. (Other anomaly detection methods~\cite{Collins:2019jip} report combined performances using the jet mass too as discriminator.) In all cases we require $\ptj \geq 1$ TeV, and we do not apply an upper cut on this variable because the distributions are in all cases concentrated towards smaller transverse momentum. 
The jet mass interval selected for {\tt hi80} taggers is $m_J \in [60,100]$ GeV, and for {\tt hi200} taggers it is $m_J \in [160,240]$ GeV. These jet mass window requirements reduce the QCD background by factors of $6.5$ and $7.1$, respectively.

\begin{figure}[htb]
\begin{center}
\begin{tabular}{ccc}
\includegraphics[height=5.2cm,clip=]{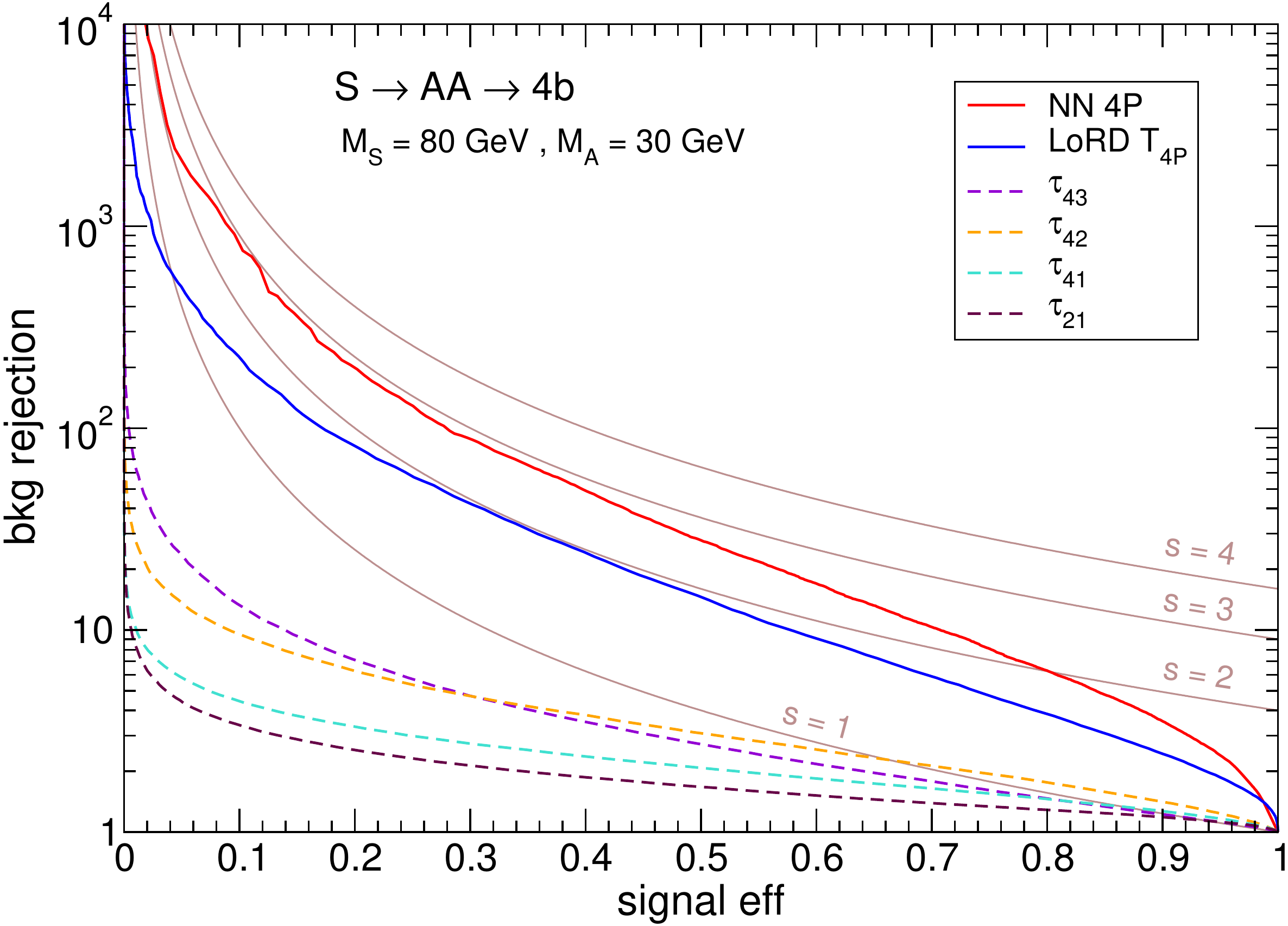} & \quad &
\includegraphics[height=5.2cm,clip=]{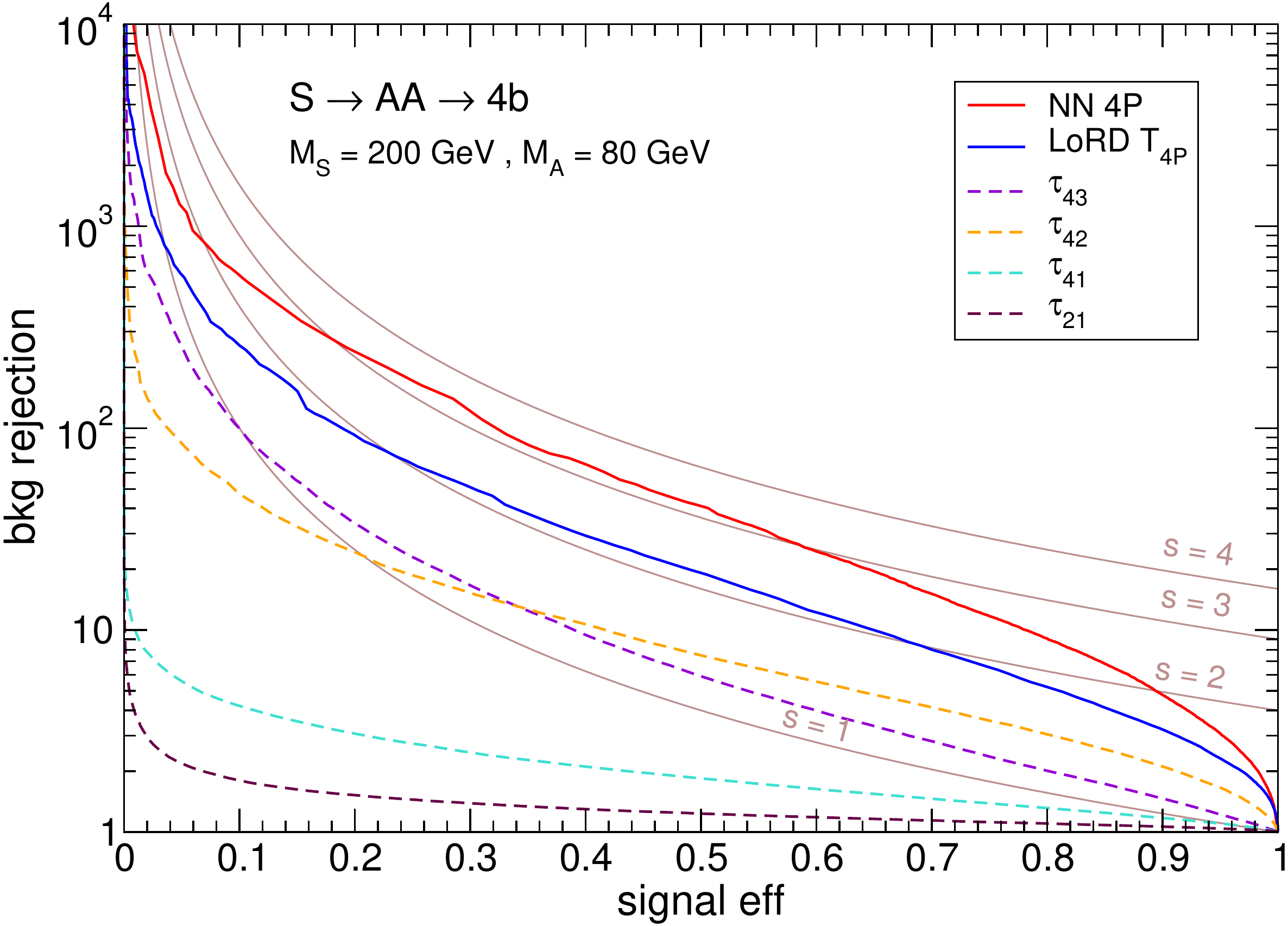} \\
\includegraphics[height=5.2cm,clip=]{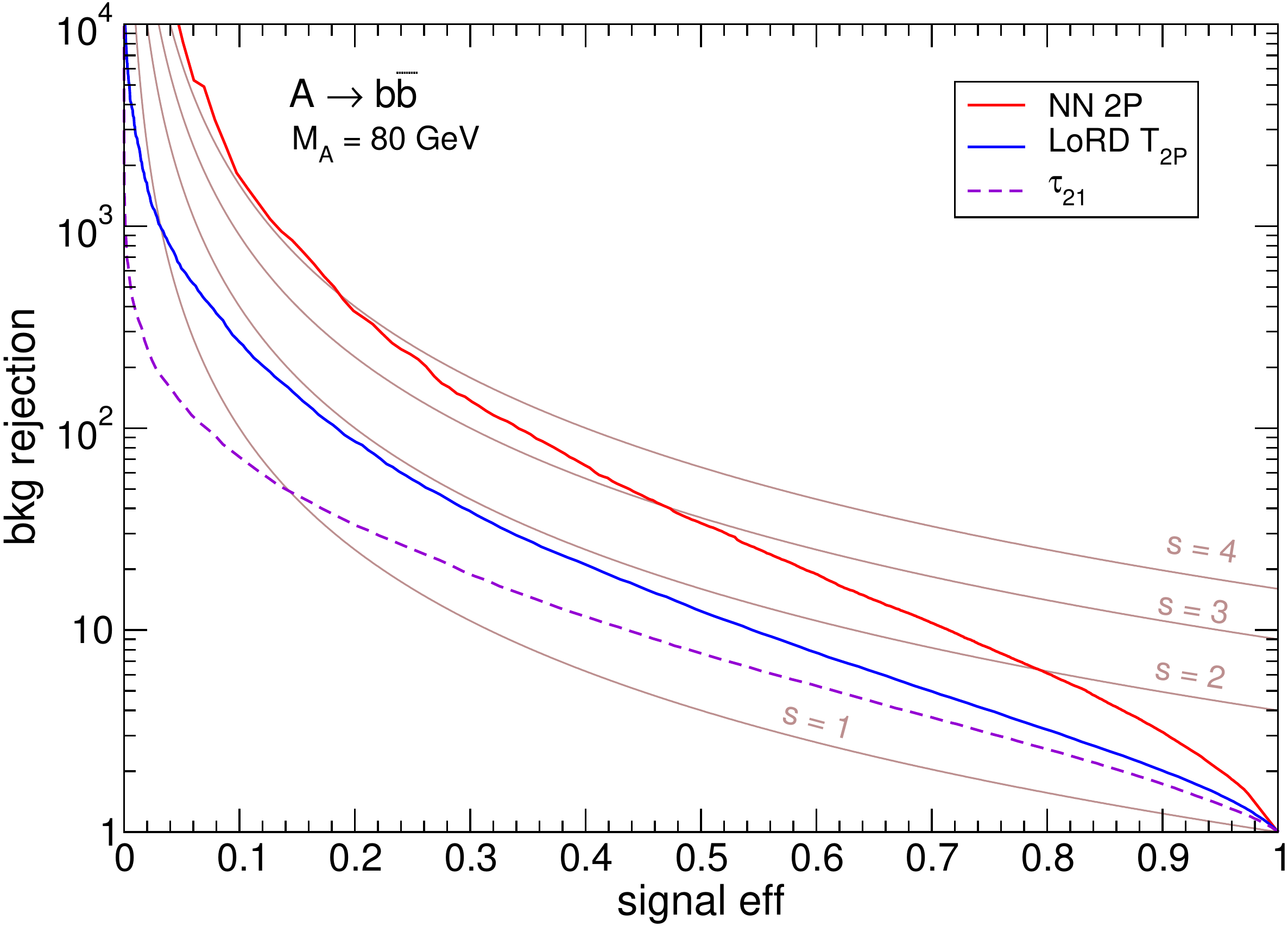} & \quad &
\includegraphics[height=5.2cm,clip=]{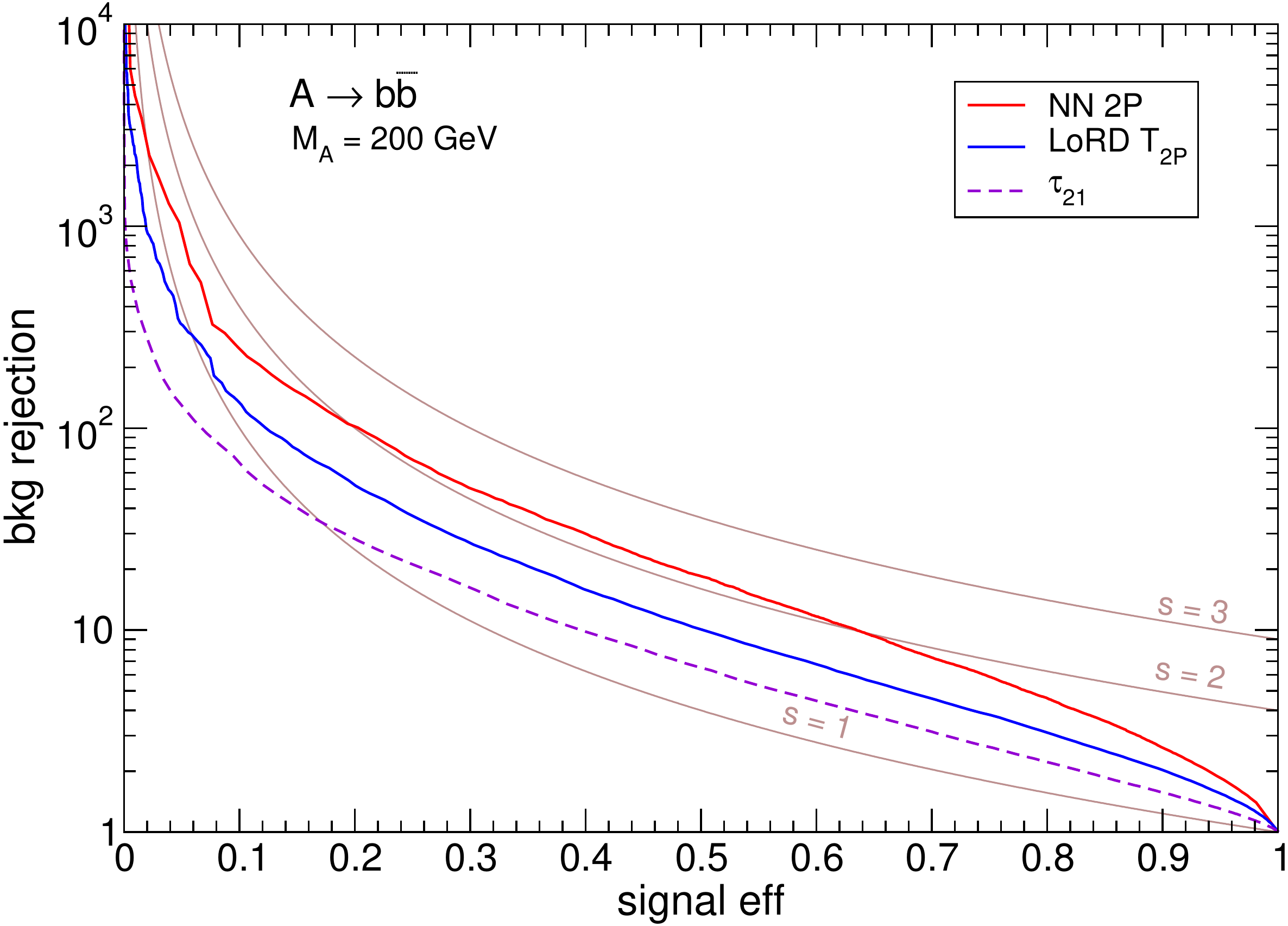} \\
\includegraphics[height=5.2cm,clip=]{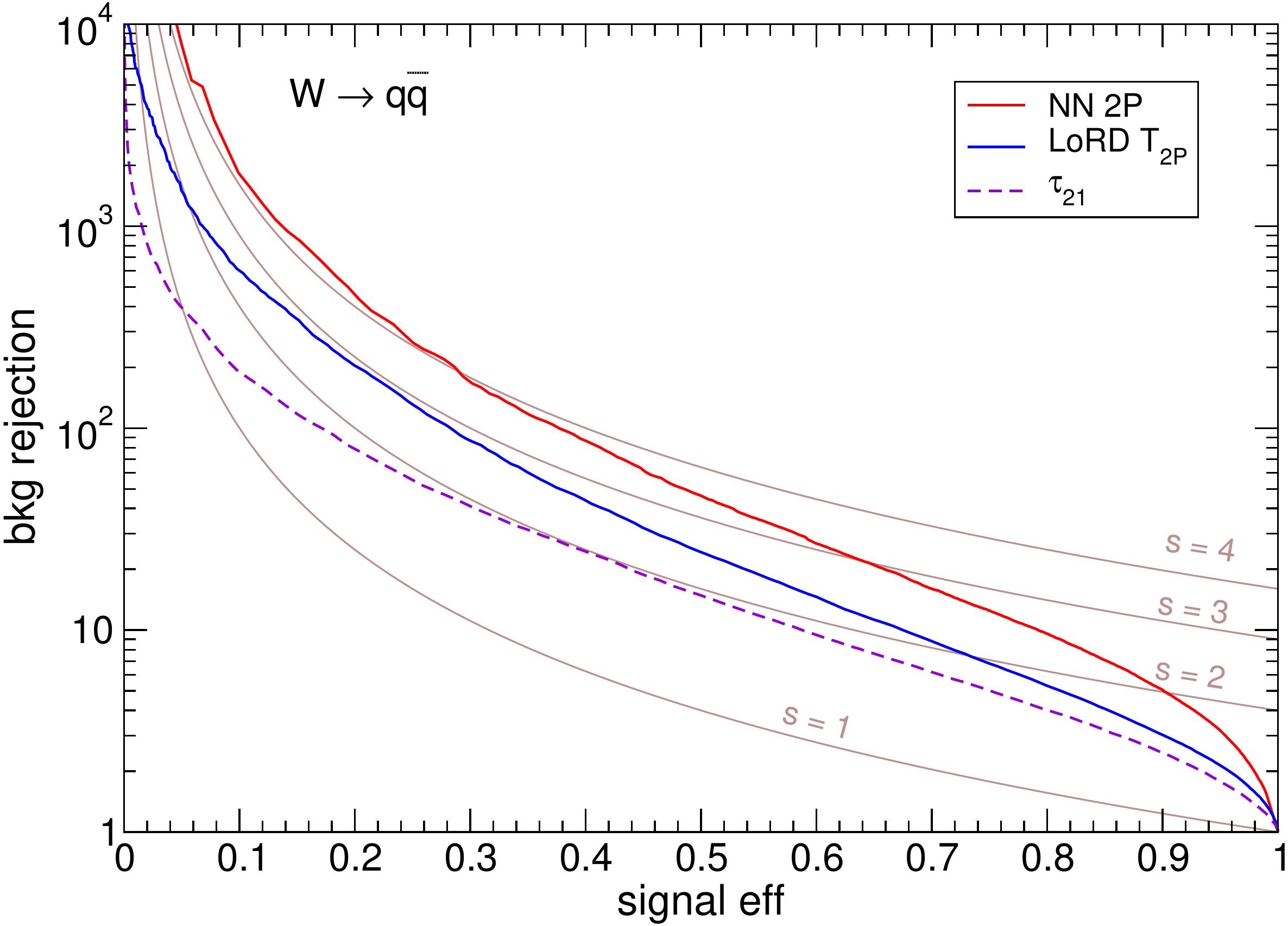} & \quad &
\includegraphics[height=5.2cm,clip=]{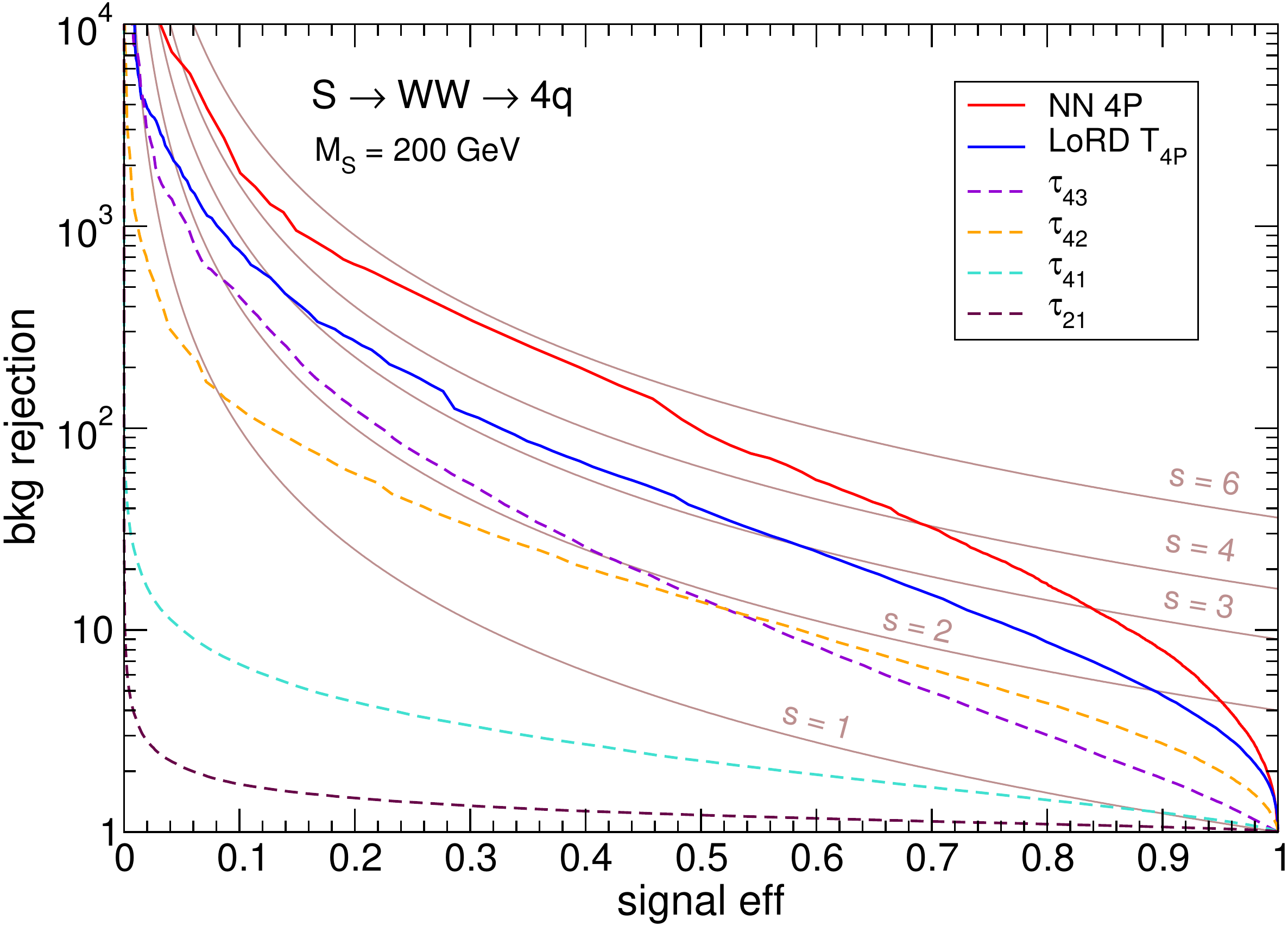} 
\end{tabular}
\caption{ROC curves for the {\tt hi80} and {\tt hi200} taggers applied on selected signals giving multi-pronged jets, compared to $\tau$-ratios and dedicated NNs (see the text).}
\label{fig:ROC}
\end{center}
\end{figure}

We present in Fig.~\ref{fig:ROC} the receiver operating characteristic (ROC) curves for signal efficiency versus background rejection of the {\tt hi80} taggers (left column) and {\tt hi200} taggers (right column), evaluated on different signals indicated in each plot. For comparison, we include the ROC curves for simple ratios $\tau_{nm} \equiv \tau_n^{(1)} / \tau_m^{(1)}$ used in the literature. We also include the ROC curves for NN taggers trained on the same (groomed) mass and $p_T$ interval, using the same architecture of Ref.~\cite{Aguilar-Saavedra:2017rzt}, with two fully connected hidden layers of 512 and 32 nodes, respectively.\footnote{We have successfully validated our implementation of the NN by reproducing the results of
Ref.~\cite{Aguilar-Saavedra:2017rzt} for the generic tagger using the same signal and background samples, and with fixed intevals of ungroomed jet mass. In that work it was shown that increasing the complexity of the NN did not improve the performance.} For the NN taggers we do not perform any mass decorrelation. The results shown here for NNs are not fully comparable to those in Ref.~\cite{Aguilar-Saavedra:2017rzt} because here we select to apply the taggers on fixed intervals of groomed mass. The substructure of QCD jets with groomed mass e.g. $m_J \in [60,100]$ GeV is not the same as for jets with ungroomed mass $m_J \in [60,100]$ GeV. This can also be noticed by comparing with the results in appendix~\ref{sec:var}, obtained for jets with ungroomed mass $m_J \in [60,100]$ GeV.

To better illustrate the effect of the tagging on the signal-to-background significance $S/\sqrt{B}$ (with $S$ standing for signal and $B$ for background) we define the significance improvement as
\begin{equation}
s = \frac{\varepsilon_S}{\sqrt{\varepsilon_B}} \,,
\end{equation}
with $\varepsilon_S$, $\varepsilon_B$ the tagger efficiencies for signal and background, respectively. This is precisely the factor multiplying the luminosity-dependent ratio $S/\sqrt{B}$ due to the tagging. We plot the lines (in gray) that correspond to several values of the significance improvement $s$. Notice that, for the mass intervals selected, an additional improvement by a factor of 2 is brought by the jet mass cut, which might even be larger for more stringent cuts on $m_J$.

The first row of Fig.~\ref{fig:ROC} shows the performance for $S \to AA \to 4b$, giving four-pronged jets with $b$ quarks. 
A scalar undergoing this type of decay has been dubbed as `stealth boson' because of its elusive nature~\cite{Aguilar-Saavedra:2017zuc}. We observe that $\tau$-ratios, especially for $M_S = 80$ GeV, fail to improve the signal significance, while the LoRD tagger can improve it by a factor of two. (For a $S \to AA \to 4u$ signal the performance is similar.) The NN tagger reaches a higher significance improvement $s=3$. 

In the second row of Fig.~\ref{fig:ROC} we study two-pronged jets from $A \to b \bar b$, which are harder to identify than $W/Z$ bosons using jet substructure. The LoRD tagger performs better than the commonly used ratio $\tau_{21}$ but, again, worse than the NN tagger.

In the third row of Fig.~\ref{fig:ROC} we examine two signals without $b$ quarks. On the left panel we have $W \to q \bar q$, for which as said the tagger $T_{2P}$ has a better discrimination than for $A \to b \bar b$ with the same mass. The performance of $T_{2P}$ is half-way between the simplest $\tau_{21}$ ratio and the more complex NN. On the right panel we show $S \to WW \to 4q$, giving a four-pronged jet with four light quarks. The performance of $T_{4P}$ on this signal is quite good (better than for the analogue with four $b$ quarks, top right panel) but it is still surpassed by the NN by a factor of two in terms of significance enhancement.

\begin{figure}[t]
\begin{center}
\begin{tabular}{ccc}
\includegraphics[height=5.1cm,clip=]{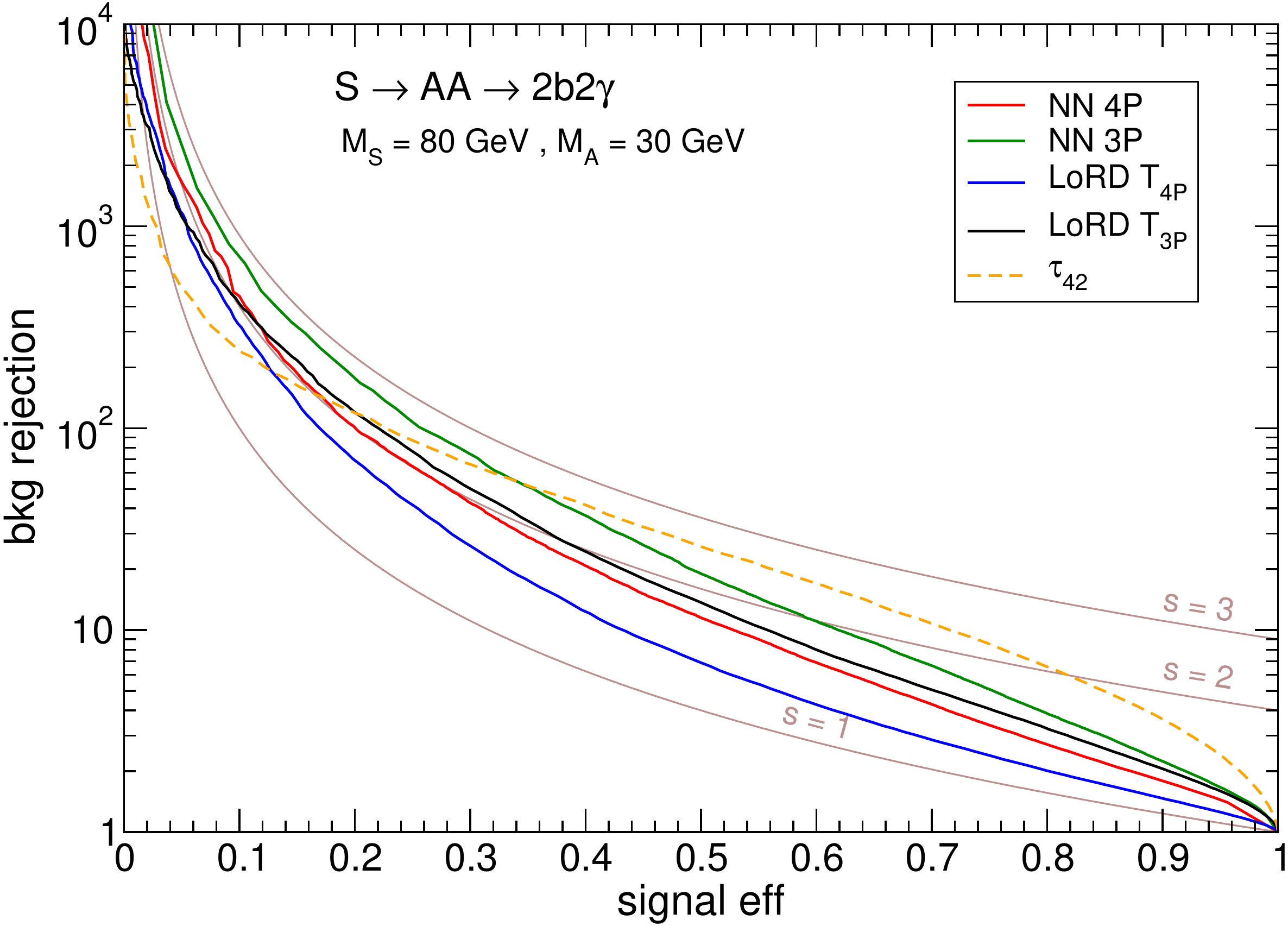} & \quad &
\includegraphics[height=5.1cm,clip=]{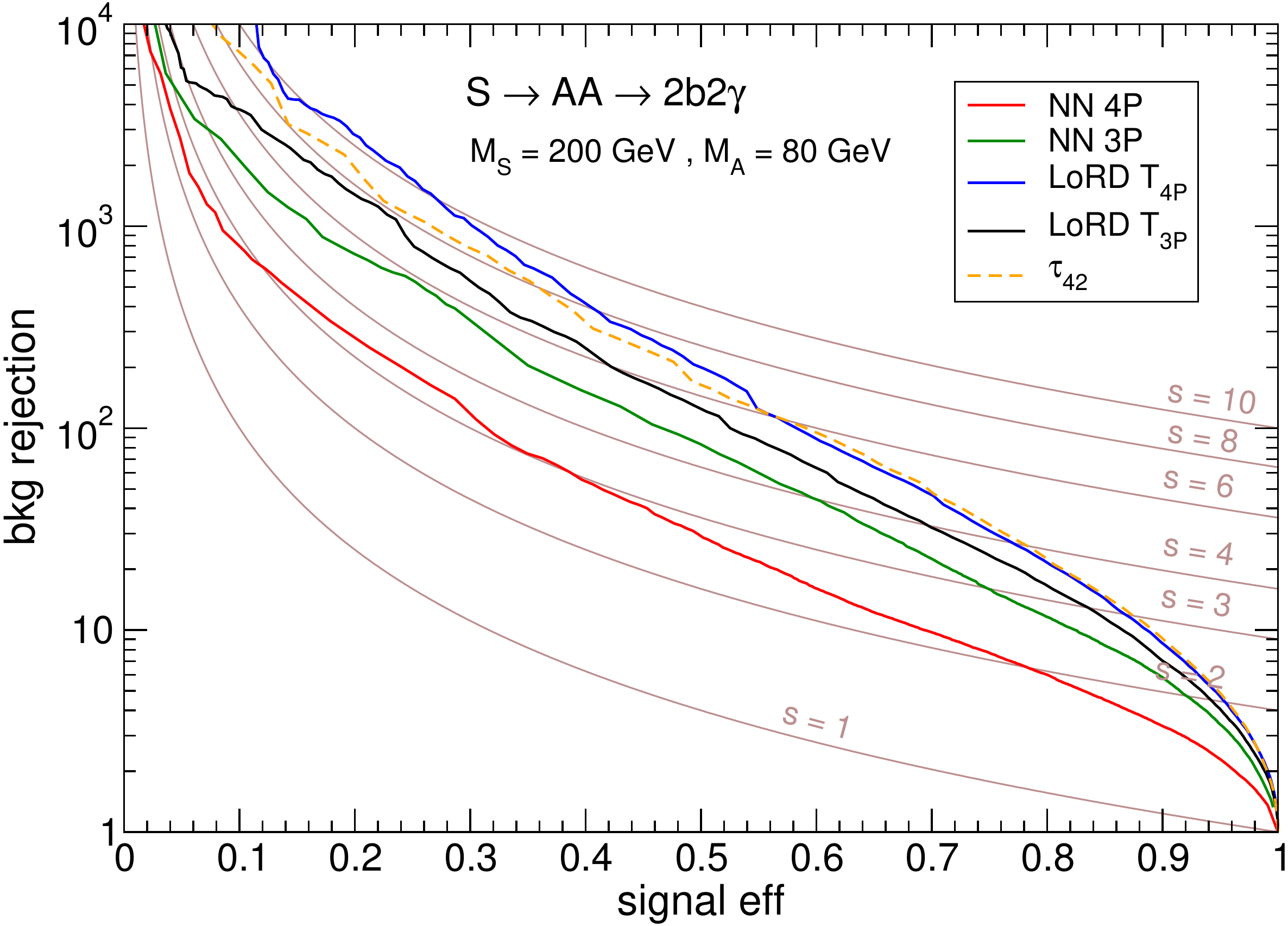} \\
\includegraphics[height=5.1cm,clip=]{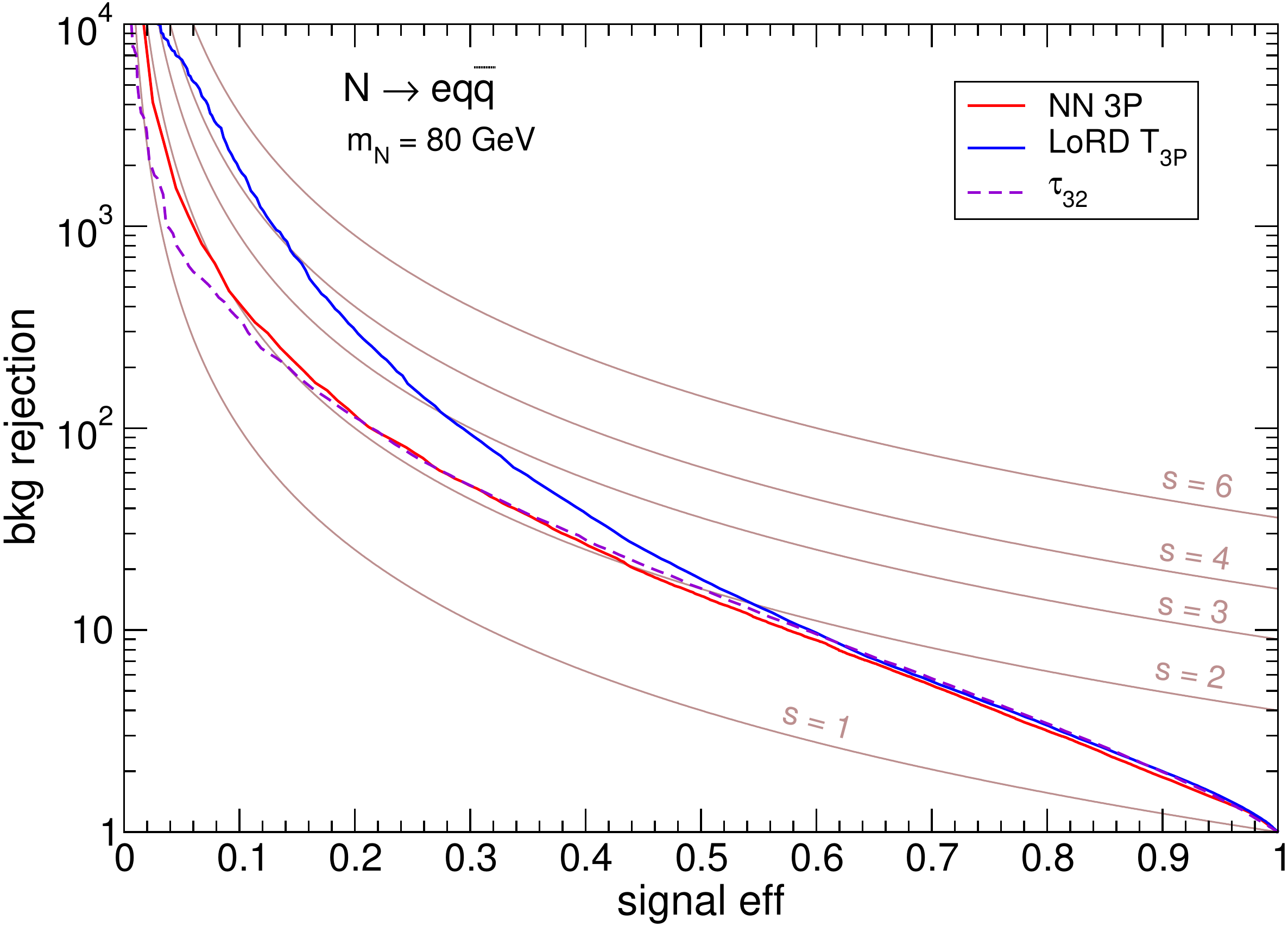} & \quad &
\includegraphics[height=5.1cm,clip=]{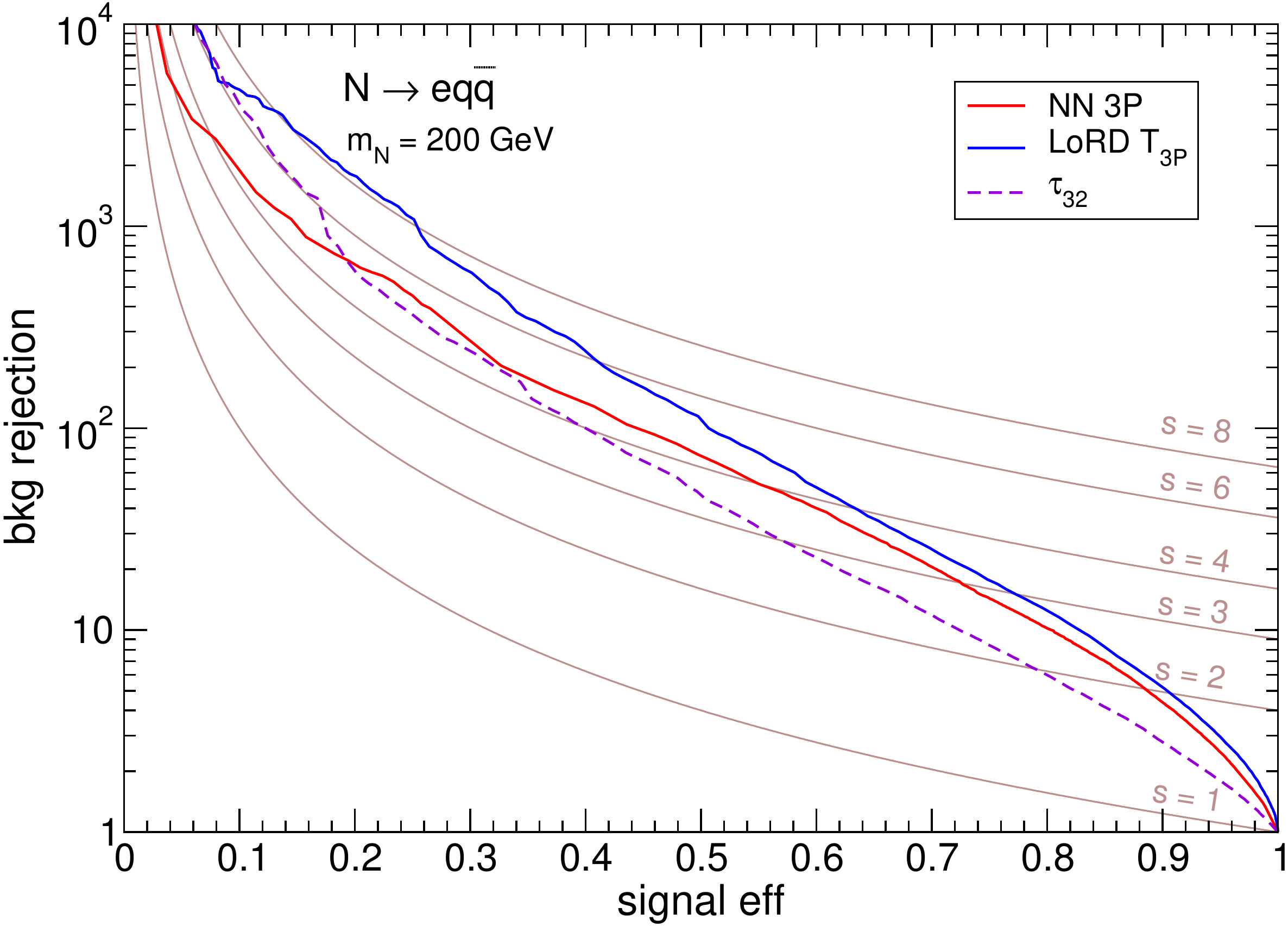}  
\end{tabular}
\caption{ROC curves for the {\tt hi80} and {\tt hi200} taggers applied on selected signals giving multi-pronged jets containing electrons or photons, compared to $\tau$-ratios and dedicated NNs (see the text).}
\label{fig:ROC1b}
\end{center}
\end{figure}

We also examine the performance of our taggers for jets containing electrons or photons as `prongs', for which the taggers are not designed. We consider
\begin{align}
& p p \to Z' \to S S \,,\quad S \to AA \to b \bar b \gamma \gamma \,, \notag \\
& p p \to Z' \to N N \,,\quad N \to e q \bar{q}  \,.
\end{align}
The decays $S \to AA \to b \bar b \gamma \gamma$ can take place for example in the model of Refs.~\cite{Aguilar-Saavedra:2019adu,AguilarSaavedra:2020wmg}.
$N$ is a heavy neutral lepton such as the right-handed neutrinos introduced in left-right models, which undergoes a three-body decay mediated by an off-shell $W_R$ boson. We set $M_{Z'} = 2.2$ TeV as before, and show our results in Fig.~\ref{fig:ROC1b}. The left column corresponds to {\tt hi80} taggers and the right column to {\tt hi200} taggers.

In the top panels of Fig.~\ref{fig:ROC1b} we show the performance of the $T_{3P}$ and $T_{4P}$ taggers for jets containing two $b$ quark and two photons. These conspicuous jets have a shape that is approximately four-pronged, and the $\tau_{42}$ ratio works well to distinguish them from QCD jets. (The ratio $\tau_{43}$ is comparable, and other $\tau$-ratios are worse; we do not show the corresponding lines for clarity.) The LoRD taggers also work well to discriminate these signals from the background. Remarkably, for higher jet masses the LoRD taggers provide a better discrimination than dedicated NNs, although their performance is similar to that of $\tau_{42}$. As it can be observed, $T_{3P}$ has a better discrimination for $M_S = 80$ GeV (top, left panel) and $T_{4P}$ is better for $M_S = 200$ GeV (top, right panel).

In the bottom panels of Fig.~\ref{fig:ROC1b} we show the performance of the $T_{3P}$ taggers  for a signal that is not properly three-pronged, since one of the `prongs' is an electron rather than a jet. The taggers  perform well for this signal for which they are not specifically designed, and better than the simple ratio $\tau_{32}$. Most surprisingly, the LoRD taggers perform better than the NN taggers, especially at $m_N = 200$ GeV. 

Overall, we find that the highest benefit of the LoRD taggers is achieved for four-pronged signals, for which they largely surpass the performance of simple $\tau$-ratios and capture a good deal of the potential of a complex NN. Also, we remark that the taggers work remarkably well for signals for which they are not designed: (a) jets containing two $b$ quarks plus two photons; (b) jets containing two light quarks plus an electron.

The performance of the LoRD taggers remains stable under moderate variations of the masses of the particle originating the fat jet. In the comparisons shown in Figs.~\ref{fig:ROC} and \ref{fig:ROC1b}, these mases were taken as 80 GeV for {\tt hi80} taggers and 200 GeV for {\tt hi200} taggers. We now investigate the results for four-pronged signals $S \to AA \to 4b$ of different masses. 
\begin{figure}[b]
\begin{center}
\begin{tabular}{ccc}
\includegraphics[height=5.2cm,clip=]{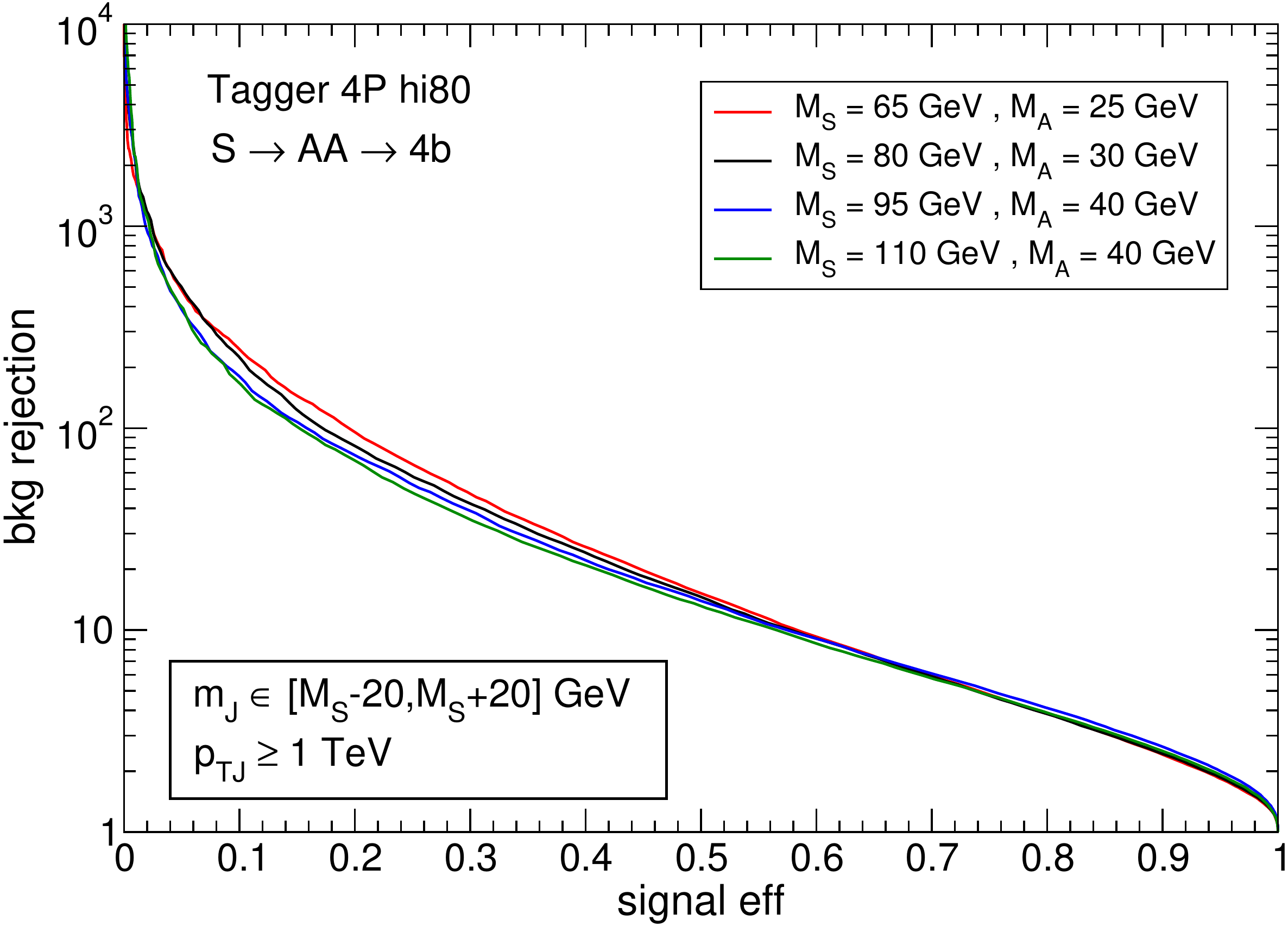} & \quad &
\includegraphics[height=5.2cm,clip=]{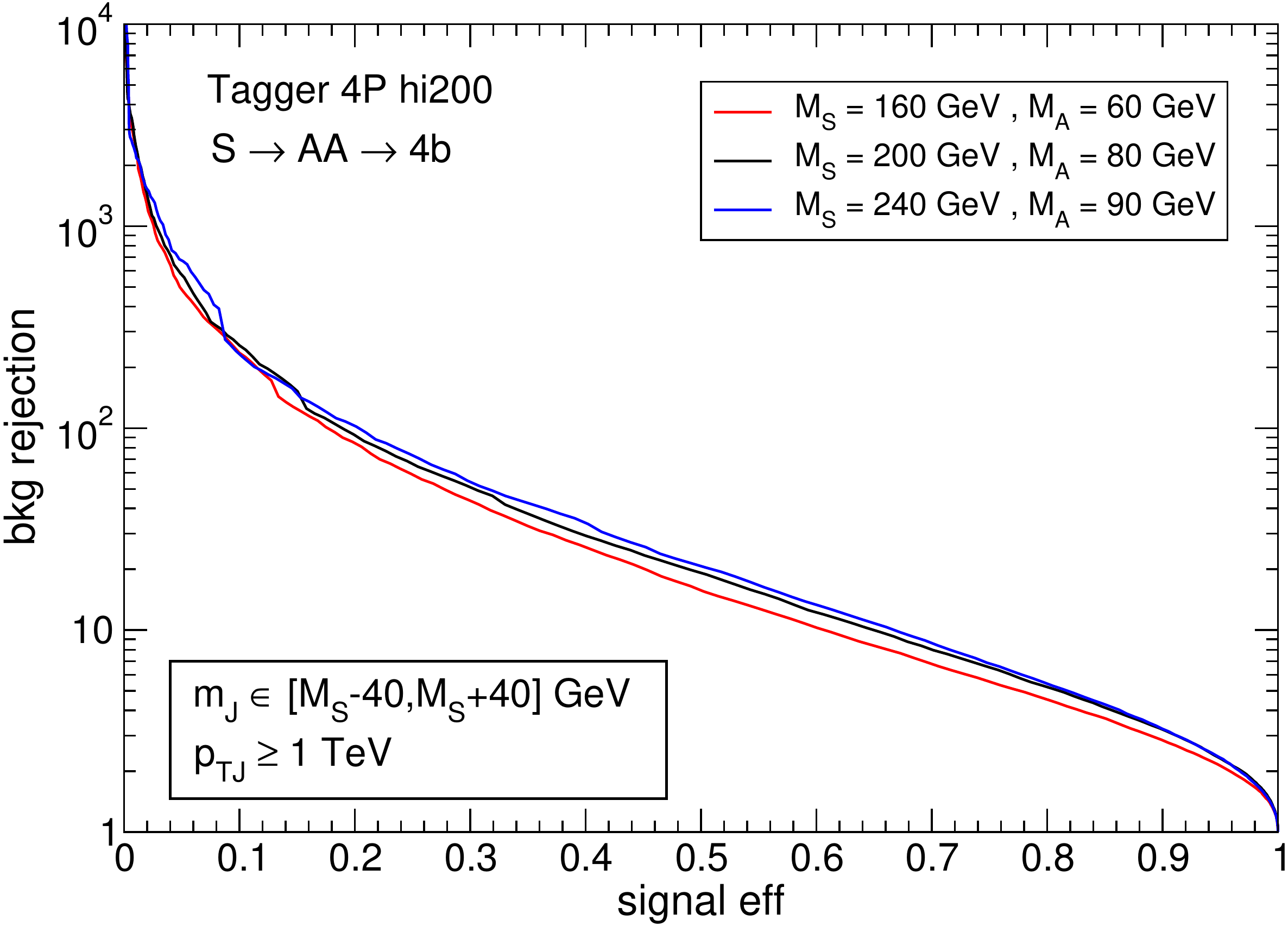} \\
\includegraphics[height=5.2cm,clip=]{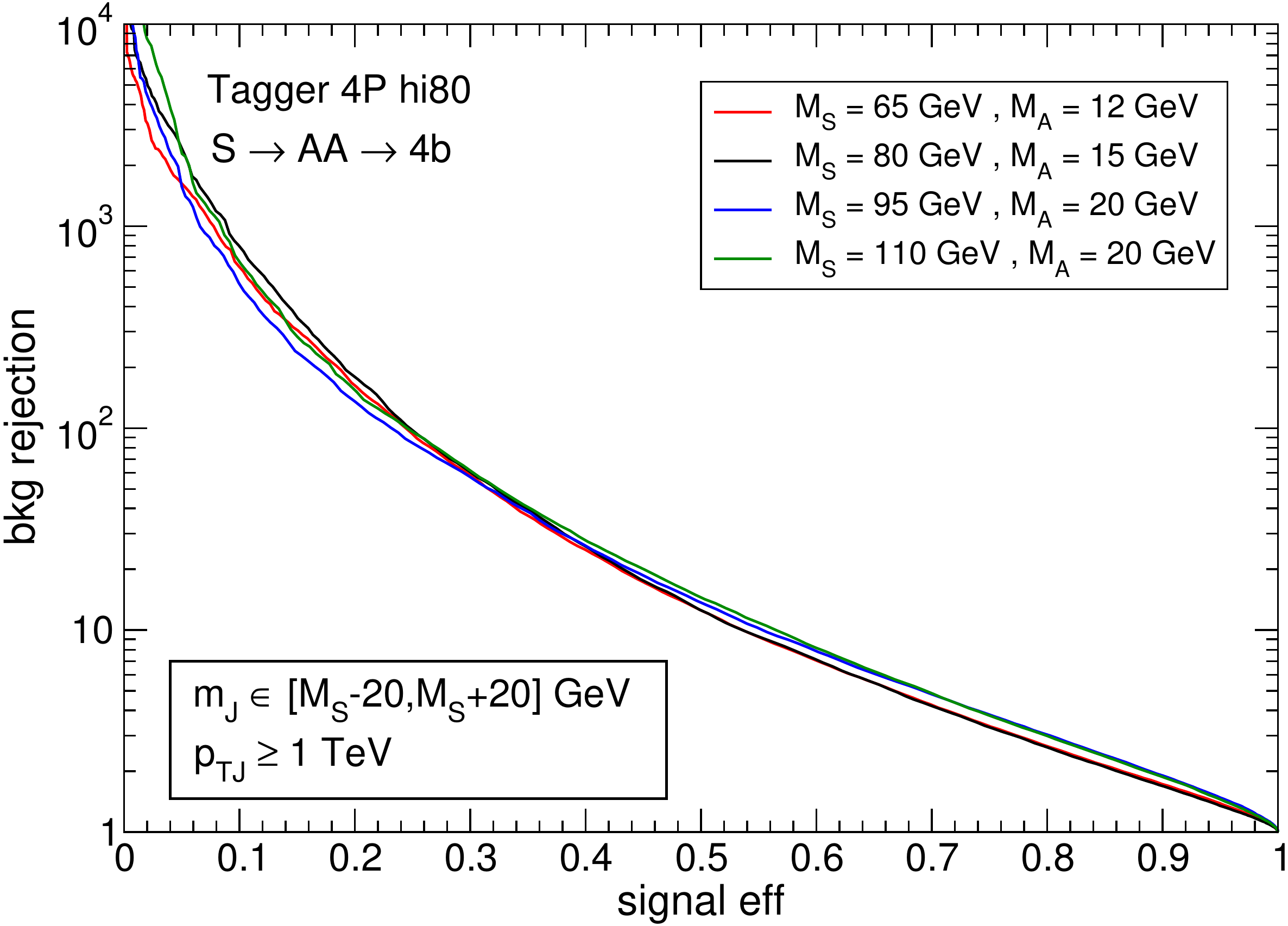} & \quad &
\includegraphics[height=5.2cm,clip=]{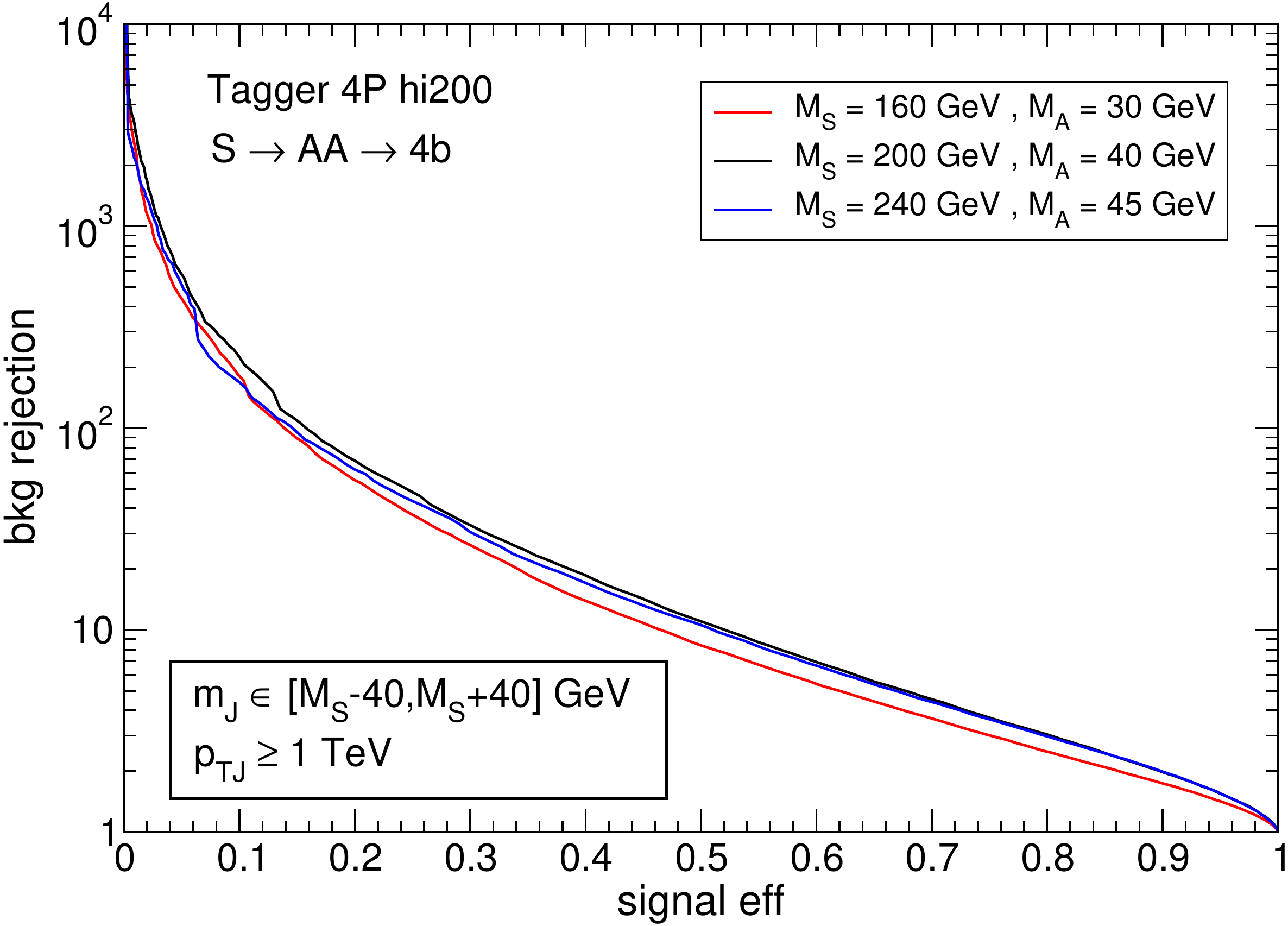} 
\end{tabular}
\caption{ROC curves for the {\tt hi80} (left) and {\tt hi200} (right) $T_{4P}$ taggers applied to $S \to AA \to 4b$ signals of different masses, see the text.}
\label{fig:ROC2}
\end{center}
\end{figure}
In the top left panel of Fig.~\ref{fig:ROC2} we show the performance of the {\tt hi80} $T_{4P}$ tagger for different masses $M_S = 65$, 80, 95 and 110 GeV. We keep the ratio $M_A : M_S \sim  30 : 80$ as in the example with $M_S = 80$ GeV, $M_A = 30$ GeV shown in Fig.~\ref{fig:ROC}, as the jet shape depends on this ratio. The jet mass cut applied is $M_S - 20~\text{GeV} \leq m_J \leq M_S + 20$ GeV. We observe that the tagger can be used for a wide range of masses, even when the mass window for the cut has little overlap with the tagger design region $[80,100]$ GeV. The bottom left panel shows the performance for the same values of $M_S$ but halving $M_A$, which makes the jet shape more two-pronged-like and increases the tagger performance.

In the top right panel of Fig. \ref{fig:ROC2} we select masses $M_S = 160$, 200 and 240 GeV with $M_A : M_S \sim 80 : 200$ and in the bottom right panel with $M_A : M_S \sim 40 : 200$. The jet mass cuts applied are
$M_S - 40~\text{GeV} \leq m_J \leq M_S + 40$ GeV.  Again, we observe that the performance of the {\tt hi200} $T_{4P}$ tagger remains quite stable under moderate variations of the jet mass.

\section{Dependence on input data kinematics}
\label{sec:4}

\begin{figure}[htb]
\begin{center}
\begin{tabular}{ccc}
\includegraphics[height=5.2cm,clip=]{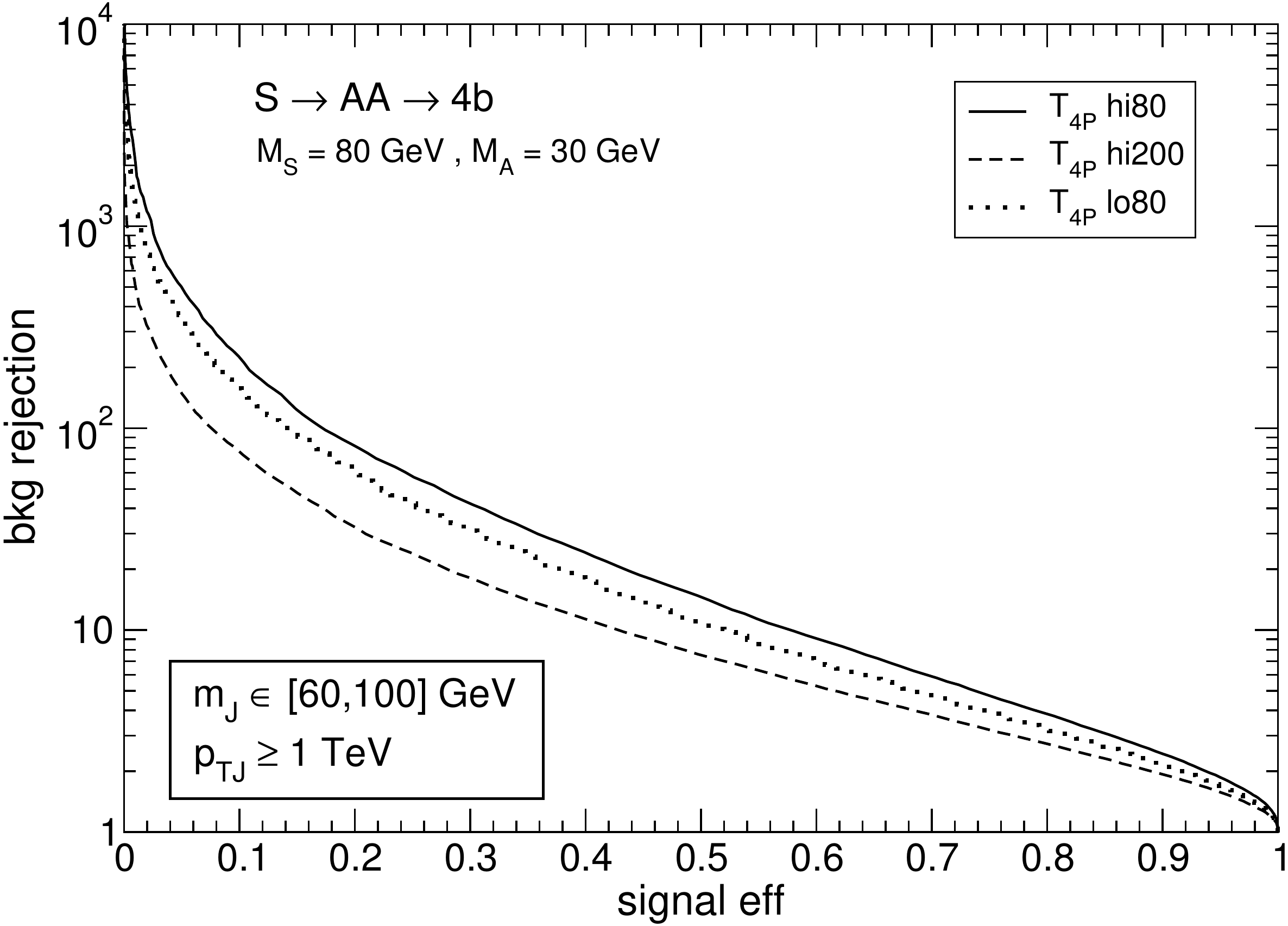} & \quad &
 \includegraphics[height=5.2cm,clip=]{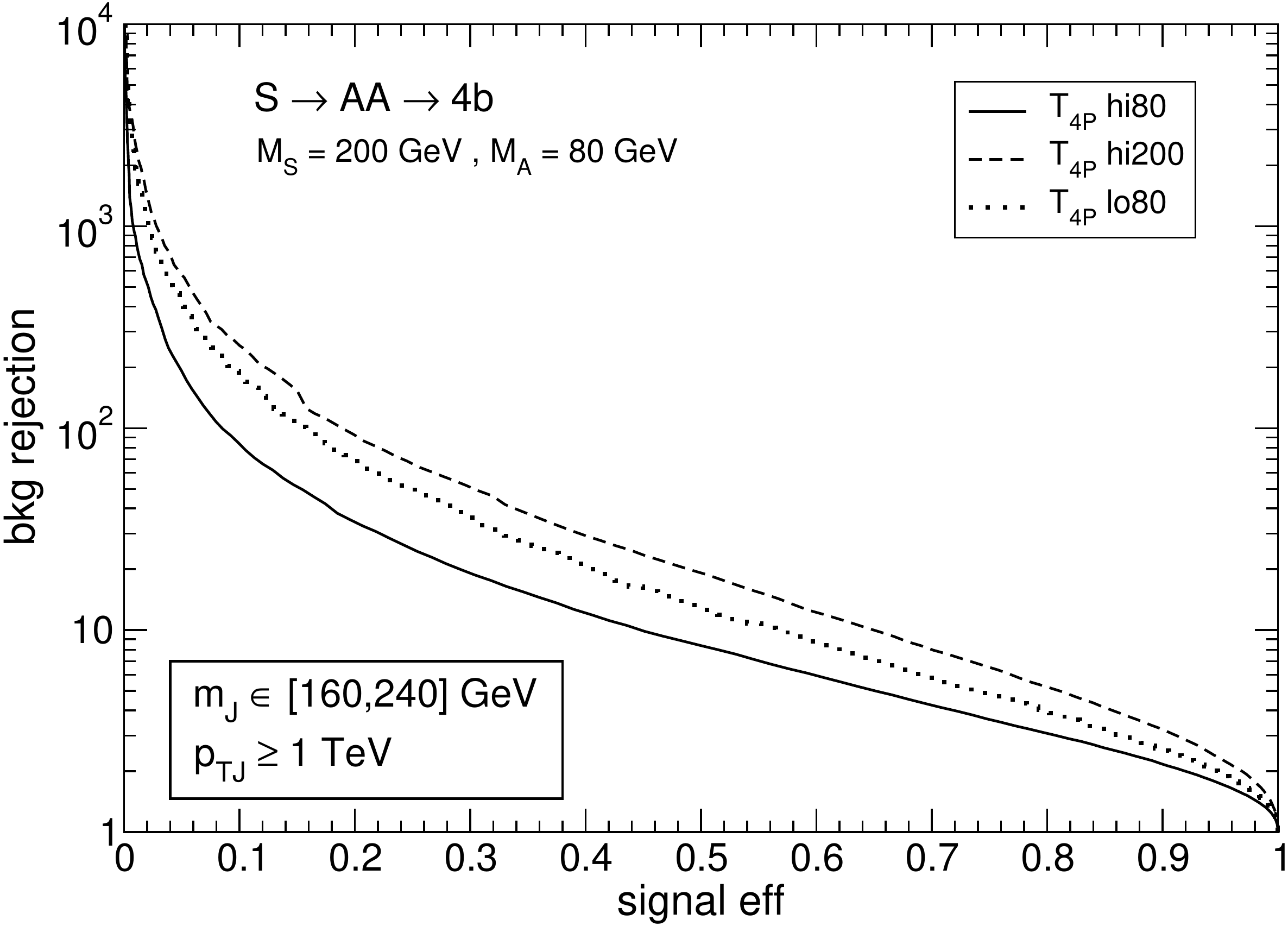} \\
\includegraphics[height=5.2cm,clip=]{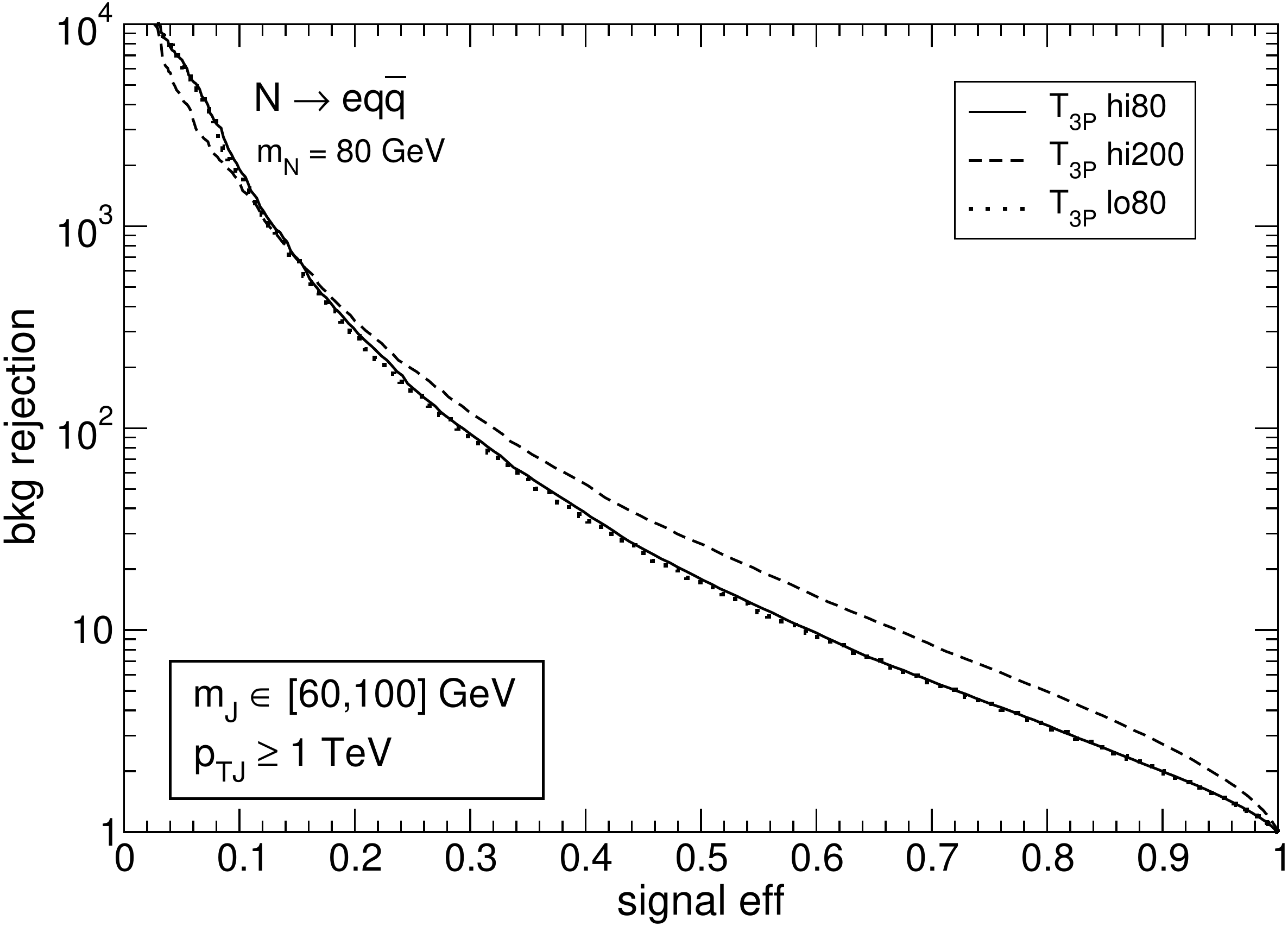} & \quad &
\includegraphics[height=5.2cm,clip=]{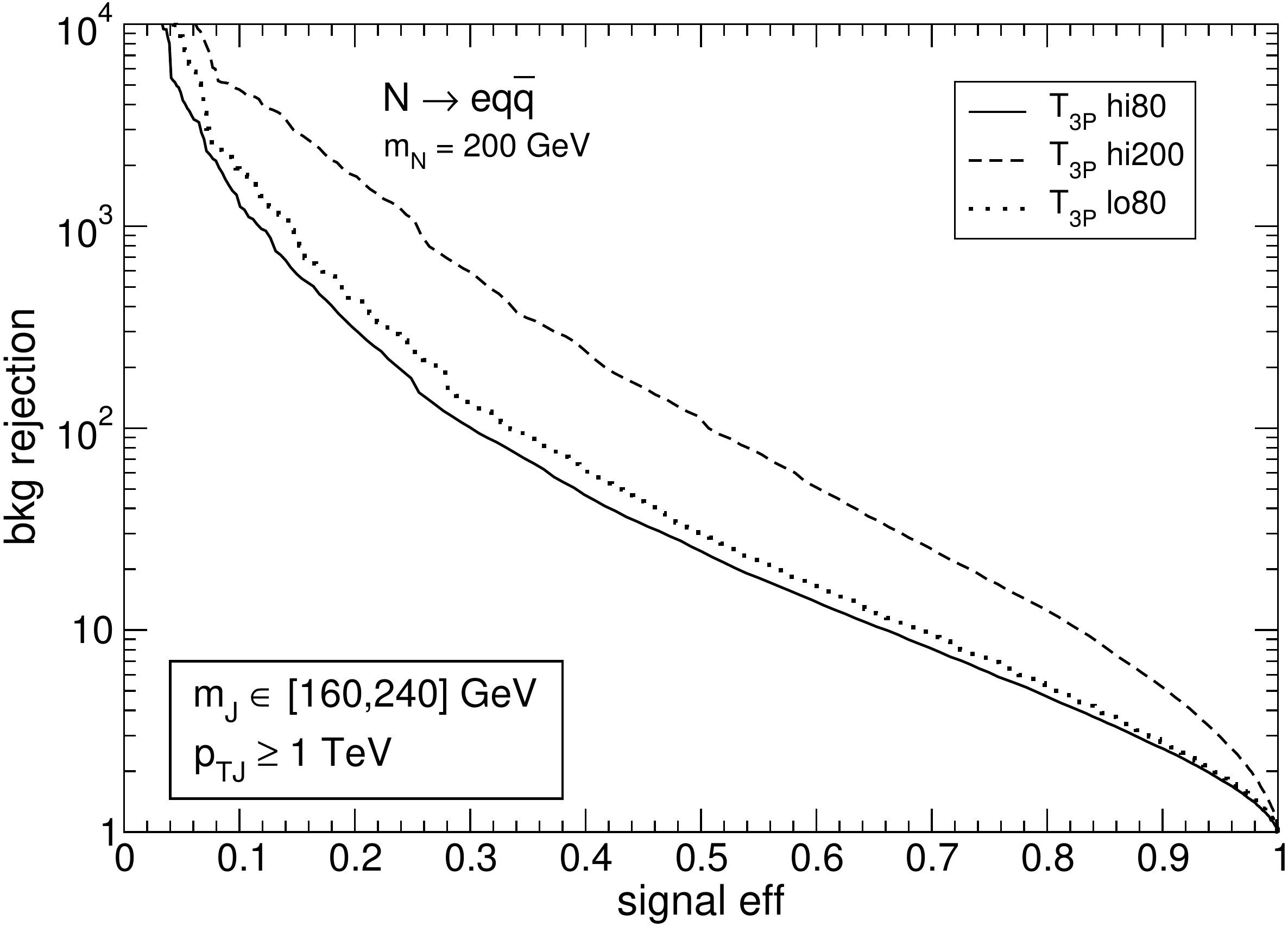} \\
\includegraphics[height=5.2cm,clip=]{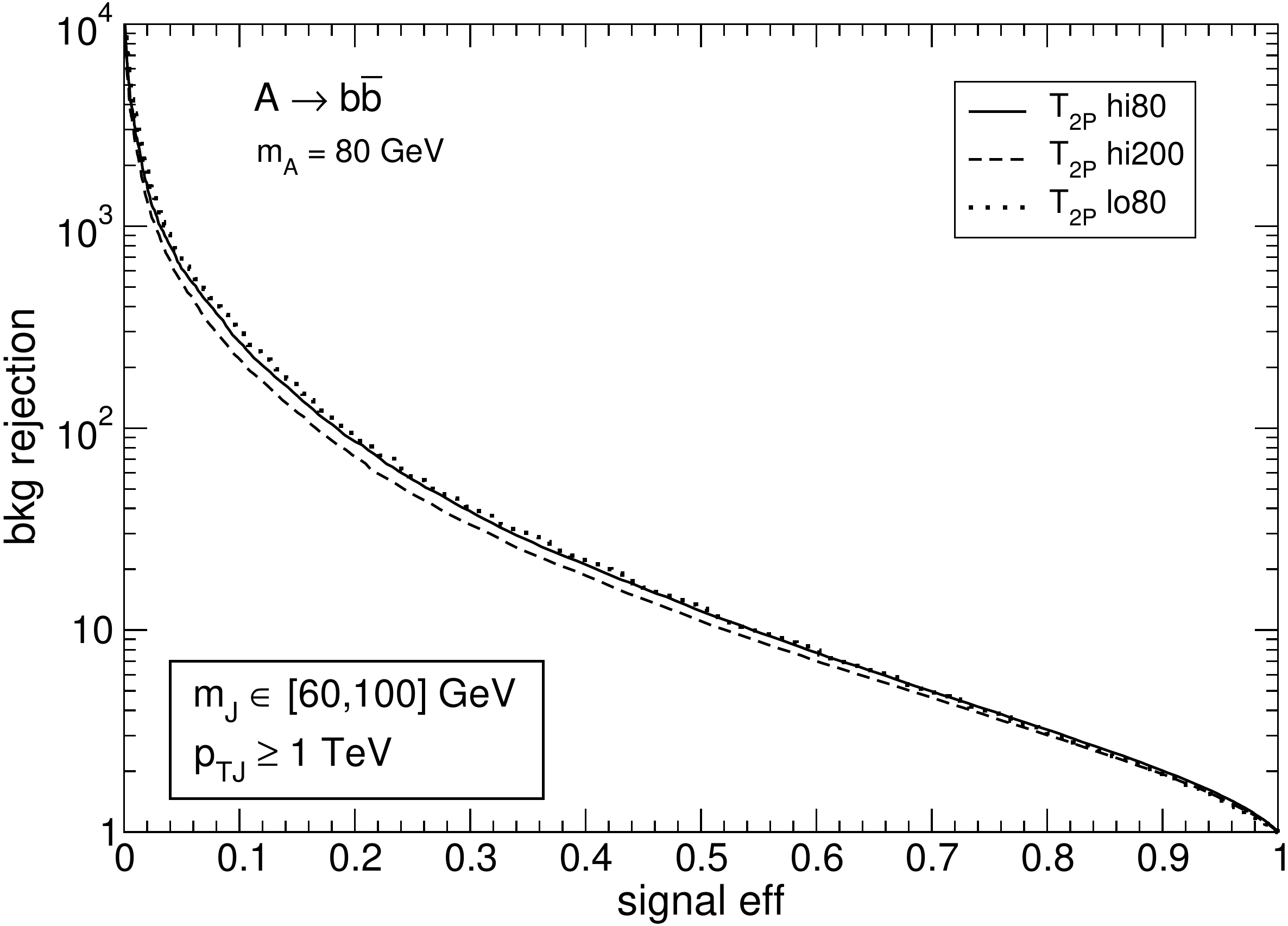} & &
\includegraphics[height=5.2cm,clip=]{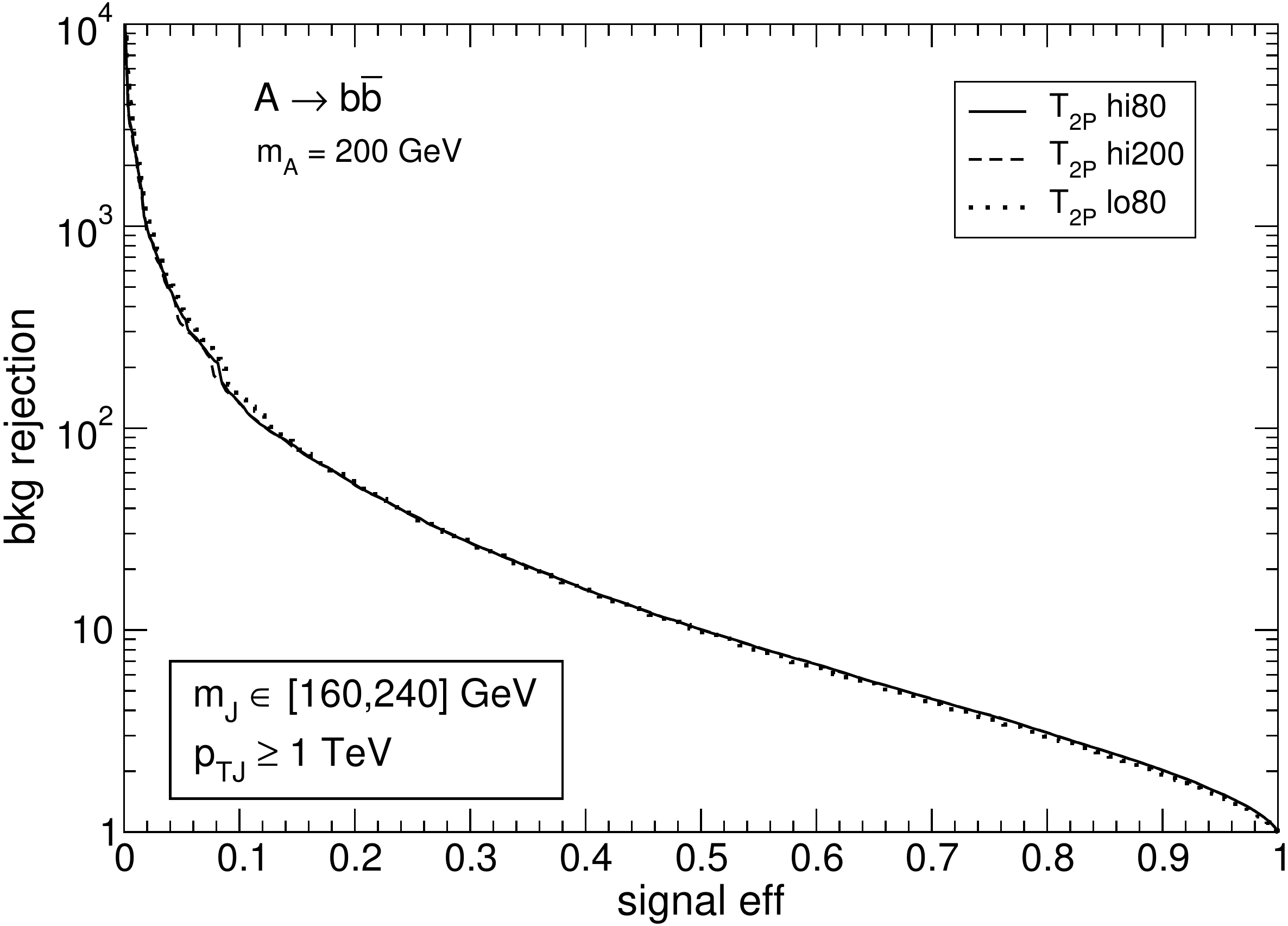}
\end{tabular}
\caption{Comparison of taggers designed on different kinematical regions ({\tt hi80}, {\tt hi200}, {\tt lo80} and applied to signals and backgrounds on reference $m_J$ and $\ptj$ intervals.}
\label{fig:compkin}
\end{center}
\end{figure}

We address in this section the variation of the tagger performance for signals (and backgrounds) with masses or transverse momentum quite different from the ones used in the design. We select the two reference mass and $p_T$ intervals used in Figs.~\ref{fig:ROC} and \ref{fig:ROC1b} of the previous section: (a) $m_J \in [60,100]$ GeV, $\ptj \geq 1$ TeV; (b) $m_J \in [160,240]$ GeV, $\ptj \geq 1$ TeV. We apply the {\tt hi80}, {\tt hi200} and {\tt lo80} taggers to selected 4P, 3P and 2P signals. The results are shown in Fig.~\ref{fig:compkin}.
The left column corresponds to the different taggers applied on signals in the kinematical region (a), and the right column to different taggers applied on signals in the region (b). From the comparison we can observe that:
\begin{itemize}
\item The performance is quite stable with $\ptj$ (solid versus dotted lines in the left column). Here we apply taggers designed with $\ptj \geq 1$ TeV ({\tt hi80}, solid lines) and $\ptj \geq 500$ GeV ({\tt lo80}, dotted lines), on jets with $\ptj \geq 1$ TeV and we observe that the performance is quite similar for 4P taggers (top panel) and basically the same for 3P and 2P taggers.
\item For 4P taggers, the tagger efficiency is degraded when applying the {\em wrong} one, i.e. the {\tt hi200} tagger in the region (a) or the {\tt hi80} tagger in the region (b). For 3P taggers, nevertheless, the {\tt hi200} tagger is good in both (a) and (b) kinematical regions. For 2P taggers there is no appreciable difference in any case.
\end{itemize}
These results suggest that, if one is interested on a very wide range of jet masses, a set of two or three LoRD taggers with different masses can be used to cover that region without performance loss. The stability of the taggers resulting from the optimisation procedure, shown in Fig.~\ref{fig:v1v2}, and the stability of the performance with small jet mass variations, shown in in Fig.~\ref{fig:ROC2}, ensure that this procedure is feasible and would lead to smooth tagging efficiencies across the whole jet mass interval.

\section{Conclusions}
\label{sec:5}

In this work we have used a logistic regression design (LoRD) to obtained simple taggers for multi-pronged jets based on jet substructure variables. These taggers can be approximately decorrelated from the jet mass. The application of the taggers keeps the shape of the jet mass distribution for the QCD background to a large extent.

The best results are achieved for taggers for four-pronged (4P) signals, which precisely are the least covered in terms of available tools. In this case, the mass decorrelation is very good (see Fig.~\ref{fig:mdecorr}), the signal to background discrimination largely surpasses simple $\tau$-ratios used in current new physics searches, and is not far from dedicated NNs. 
For boosted top quarks, the three-pronged (3P) taggers do not bring much improvement over the ratio $\tau_{32}$.\footnote{This result may be due to the fact that the 3P taggers are designed by using three-pronged jets with $udd$ and $ubb$ content in a colour singlet, while the top quark contains $u \bar d b$ or $c \bar s b$, in a colour triplet. This possible effect deserves further investigation.}
However, for boosted heavy neutrinos --- which give jets that are not properly three-pronged --- the LoRD taggers perform very well, even better than NNs trained on the same set of signals and backgrounds. A significance improvement by a factor up to 8 can be achieved. This fact is quite interesting since current searches for boosted heavy neutrinos \cite{Aaboud:2019wfg} do not use any type of jet substructure analysis. 
We have also tested two-pronged (2P) taggers on a variety of signals, finding that the performance is half-way between the ratio $\tau_{21}$ and a dedicated NN. LoRD $T_{3P}$ and $T_{4P}$ taggers are also sensitive to jets containing two $b$ quarks plus two photons, a signature which is not experimentally covered~\cite{AguilarSaavedra:2020wmg}.

We envisage two possible situations where the LoRD taggers may be very useful. The first one is when the development of a full-fledged multivariate tagger with mass decorrelation is not feasible. In this case, a LoRD tagger (or a handful of them) can easily be used, obtaining results that are not far from the ones that a multivariate method could bring.

The second situation is as a complement and cross-check of results obtained with more complex taggers as the ones based on deep neural networks. These methods are often a `black box' whose results are difficult to test independently. Because the performance of the LoRD taggers is not far from NNs, they can be very useful as a robust test, especially in case any new physics signal involving fat jets is found at the LHC.

\section*{Acknowledgements} We would like to thank A. Casas, J. R. Espinosa and J. M. No for useful discussions. This work has been supported by Spanish Agencia Estatal de Investigaci\'on through the grant `IFT Centro de Excelencia Severo Ochoa SEV-2016-0597' and by MINECO project FPA 2013-47836-C3-2-P (including ERDF). B.Z. is further supported by the Programa Atracci\'on de Talento de la Comunidad de Madrid under grant n. 2017-T2/TIC-5455,
and from the Comunidad de Madrid/UAM project SI1/PJI/2019-00294.

\appendix

\section{Logistic Regression implementation}
\label{app:LR}
In this appendix we describe in detail our implementation of the Logistic Regression classifier for distinguishing signal from background.

The datasets consist of $N$ pairs $\{{\bf x}_i , t_i\}_{i=1}^N$, where $t_i$ are the binary labels of each event (signal or background) and ${\bf x}_i$ are the input variables, in our concrete case ${\bf x}=\log \;{\boldsymbol\tau}$, where ${\boldsymbol\tau}$ is the vector of subjettiness variables as defined in (\ref{ec:taulist}). We preprocess the data such that signal and background events are equal in number for the training set and validation sets. For the latter, we use 20\% of the dataset, while the remaining 80\% is used for training (note that the test data in our study are completely independent samples analysed a posteriori).   

The Logistic Regression is a simple ML model for classification where the different classes of data points are assumed to be linearly separable in the space of input variables ${\bf x}$. As a probabilistic discriminative model, its goal is to model the conditional probability $p(C_k|{\bf x}_i)$ of a given class $C_k$, with $(k=0,1)$, given an input ${\bf x}_i$, and it does so with the following parametric form:
\begin{equation}
p(C_0  |{\bf x}_i) \equiv y_i ({\bf w}) = \sigma({\bf x}_i^\top {\bf w}+w_0)
\label{ec:LR}
\end{equation} 
where we identify $C_0$ as the signal class. In the above equation, $\sigma$ is the sigmoid function, $\sigma(z)=(1+e^{-z})^{-1}$, and ${\bf w}, w_0$ are the weights to be optimised. In our analysis we have dropped the `intercept' $w_0$, since we wanted to make a direct correspondence with the product defined in (\ref{ec_Tbarprod}), which has shown to be a good signal-vs-background discriminator in the literature~\cite{Datta:2017lxt} for simple jet topologies. A non-zero intercept amounts to a global multiplicative constant in (\ref{ec_Tbarprod}). We have checked in any case that the inclusion of the intercept does not have a visible impact in the classification performance. On the other hand, the weights ${\bf w}$ are identified as the coefficients $c^\beta_n$ in Eq.~(\ref{ec:Tbar}). The optimisation of the ${\bf w}$ parameters has been done as usual by minimising the so-called cross-entropy cost function:
\begin{equation}
{\cal C}({\bf w}) = -\sum_{i=1}^N \big[ t_i {\rm ln} y_i({\bf w}) + (1-t_i){\rm ln}(1-y_i({\bf w})) \big]~.
\end{equation}   
The implementation of the above procedure has been performed with the automatic differentiation tool {\scshape Tensorflow v1.14}~\cite{tensorflow2015}. We have used the {\scshape Adam} stochastic gradient descent optimiser with learning rate $0.01$, and otherwise default parameters, and using a mini-batch of size $n=200$. Convergence is typically achieved at around 200 epochs, the training process taking around 3 seconds in an Intel(R) Core i5-5200U CPU @ 2.20GHz with 8GB of RAM.

It is worth noting that as a result of the above optimisation method, different random initialisations\footnote{We use the so-called `Xavier' initialization for the weights.} of the weights ${\bf w}$ will lead to optimal weights having ${\cal O}(1)$ differences. Even if such differences will not have an important impact in the error rate of the validation test, they will lead to slightly different distributions for the background and the signal when applying the trained model to the test samples. An example of these variations is shown in Fig.~\ref{fig:v1v2}. 

\section{Estimating the variance of the predictions}
\label{sec:var}

Apart from the variability in the performance due to the stochastic optimisation procedure described in appendix \ref{app:LR}, one may wonder about the variability of the predictions associated to the statistical variance of the parameters ${\bf w}$ that best fit the data. A clear way to proceed would be to adopt a Bayesian approach, where a proposed prior distribution for the parameters is updated with data (via the likelihood) in order to obtain the posterior distribution, which can be used in a second step to obtain the probability distribution of the prediction for a given new input. However, even if for the ML model in consideration (Logistic Regression) this would be a tractable task, the procedure would still require the use of approximate inference techniques in order to approximate the posterior distribution mentioned above, as well as to approximate the predictive distribution. 

Instead, we opt here for a less ambitious frequentist alternative to estimate the variance of the parameters, known in some literature as the bootstrapping method. The idea is very simple: (i) generate a set of $B$ synthetic datasets identical in size to the original dataset; (ii) train the ML model on each of those, such as to have $B$ sets of optimal parameters; (iii) from the previous step build the resulting distribution of the parameters, from which to obtain the desired variance. 

In the procedure outlined above, a decision has to be made about how to generate the synthetic datasets. Ideally, one has at hand a `generative model' (for example, the likelihood of the data) which it is possible to sample from by using the optimal parameters obtained with the original dataset. However, in this work we have used a discriminative model (see appendix \ref{app:LR}), where the likelihood of the data is not modeled. Then, an alternative --- followed in this work --- is to sample directly from the original dataset `with replacement', as it is customary in the bootstrapping literature. In other words, we generate $B=50$ datasets by randomly choosing points from the original dataset, such that some of those points may be present more than once in the synthetic datasets. 

\begin{figure}[htb]
\begin{center}
\includegraphics[height=10cm]{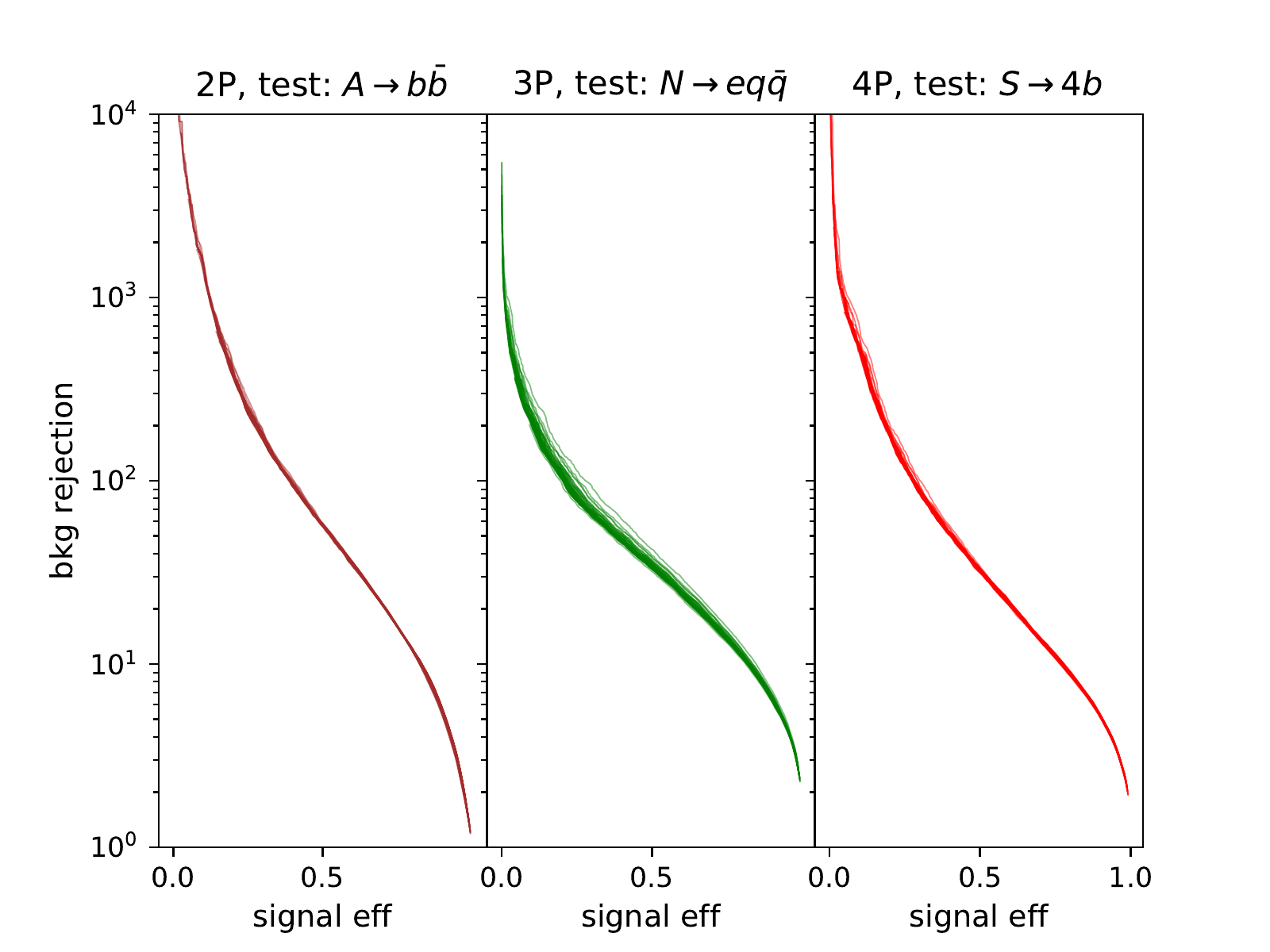} 

\caption{Variability in the performance of 2P, 3P and 4P taggers (from left to right, respectively), obtained from $B=50$ bootstrap samples of the original dataset.}
\label{fig:bootstrap}
\end{center}
\end{figure}

Finally, as a shortcut to computing the predictive distribution, we proceed directly to evaluate the performance of the $B$ set of fitted parameters under the test dataset. The results are illustrated in Fig.~\ref{fig:bootstrap} for the {\tt hi80} kinematical region, and for test samples corresponding to an ungroomed mass interval $m_J\in [60, 100]$ GeV. 

 We observe that the variability in the predictions is very small for the three taggers, which is a desirable outcome in any statistical modeling. We note that the estimation of such variability may be underestimated at some extent due to the finiteness of the original dataset, which may bias the bootstrap samples.    

\section{Interpretability of the results and dimensionality of the datasets}
\label{sec:PCA}
One of the main motivations of this work is to obtain simple enough (yet efficient)  taggers, providing the interpretability of the results which is typically absent in more complex multivariate models as neural networks. Interpretability is intended here first of all in the sense of the parametric form of the decision boundary between signal and background, and clearly the model adopted in this work, Logistic Regression, featuring a linear decision boundary in the input space, is more interpretable than more complex models.

A different, more physical sense of interpretability could be explored, regarding the input variables themselves. The basis of subjettiness variables (\ref{ec:taulist}) is complete in the sense that it completely specifies the variables of $M$-body phase space (being $3M-4$ in total) such as the transverse momenta, relative angles, etc. up to a global rotation. Consequently, it is expected that {\it all} the subjettiness variables play an important role in the signal/background discrimination for $M$-pronged jets.

Still, in this work we have used values of $M$ that are higher than the number of prongs of the `signal' jets, e.g. we have used $M=9$ to build taggers for four-pronged jets. Therefore, one may wonder whether the full set of variables is required. We have verified that this is the case, by building 4P taggers where, instead of using all the $\tau_n^{(\beta)}$ variables of the basis (\ref{ec:taulist}), we restrict ourselves only to the subsets with either $\beta=1$, $\beta=2$ or $\beta=0.5$. The performance of the resulting taggers are significantly worse for the {\tt hi80} test samples, having for a signal efficiency of 0.2 a corresponding background rejection an ${\cal O}(10)$ smaller than the 4P tagger with the full set of variables (cf. Fig.~\ref{fig:ROC1b}). 

Note, however, that the fact that the full set of variables is required for optimal signal to background discriminetion does not mean that the intrinsic dimensionality of the data is $3M-4$. There is a partial correlation among the different $\tau_n^{(\beta)}$ variables and thus the dataset is expected to lie in a smaller dimensional space. In order to check this we have performed a simple Principal Component Analysis (PCA) on the original {\tt hi80} training sample. The results show that the first two principal components (i.e. the two most important rotated variables) already contain around $86\%$ of the original variance. The weights that each of the original variables have on those principal components are all similar, of ${\cal O}(0.1)$. Finally, if using only those two principal components as new input for the LoRD model (i.e. now having a 2-dimensional dataset) we obtain a performance on the {\tt hi80} samples which is very similar to the one obtained with the original $(3M-4)$-dimensional dataset. This confirms our expectations that the information of the dataset is (a) contained in a hyperplane of (much) smaller dimensionality, while (b) being spread over all the $3M-4$ original variables. Further exploration of the interpretability of this kind of tagger is left for future work.

\section{Using groomed versus ungroomed quantities}
\label{sec:newa}

\begin{figure}[htb]
\begin{center}
\includegraphics[height=5.2cm,clip=]{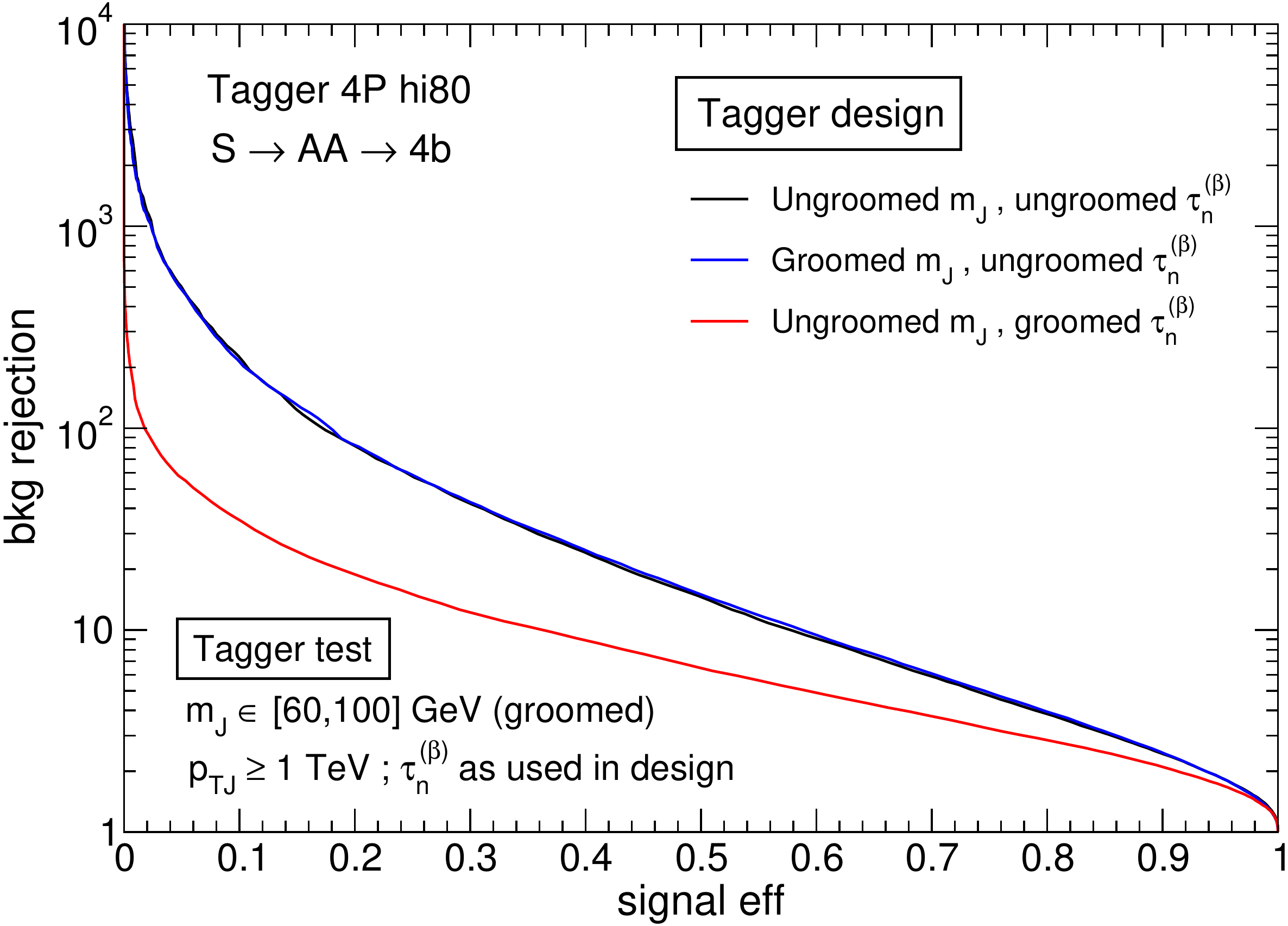} 

\caption{Comparison of the performance of the 4P {\tt hi80} tagger using groomed versus ungroomed quantities.}
\label{fig:groomvsnot}
\end{center}
\end{figure}

As mentioned in Section~\ref{sec:2}, in the design of the taggers we use MI data as signal and QCD samples as background, within fixed intervals of ungroomed mass, and for $p_T$ above and around some fixed cut. In the test of the taggers, however, we use the groomed mass and $p_T$, as it is usually done in the LHC experiments. We have found that the performance when using groomed quantities in the design is rather similar, as shown for example for the 4P {\tt hi80} tagger in Fig.~\ref{fig:groomvsnot}. The black line represents our default choice, whereas the blue line corresponds to designing the tagger using groomed $m_J$ and $p_T$, in the same intervals. On the other hand, using groomed subjettiness variables $\tau_n^{(\beta)}$, both in the design and the test, significantly degrades the performance of the tagger (red line). The use of ungroomed $\tau_n^{(\beta)}$ is not a problem, actually it is the standard choice for the CMS Collaboration, which uses alternative methods such as PUPPI~\cite{Bertolini:2014bba} to remove pile-up.

\section{Taggers without mass decorrelation}
\label{sec:a}

\begin{figure}[htb]
\begin{center}
\begin{tabular}{ccc}
\includegraphics[height=5.2cm,clip=]{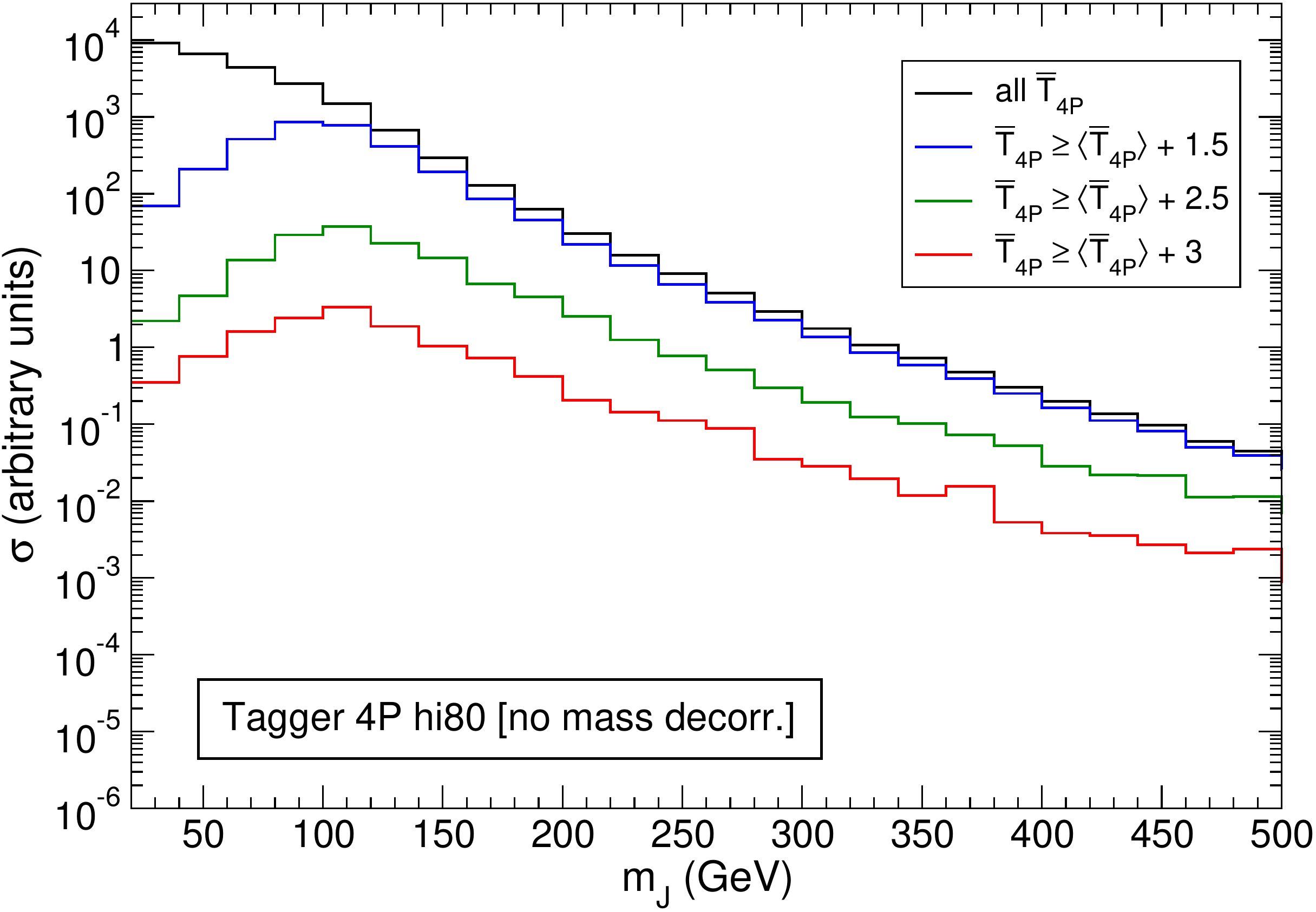} & \quad &
\includegraphics[height=5.2cm,clip=]{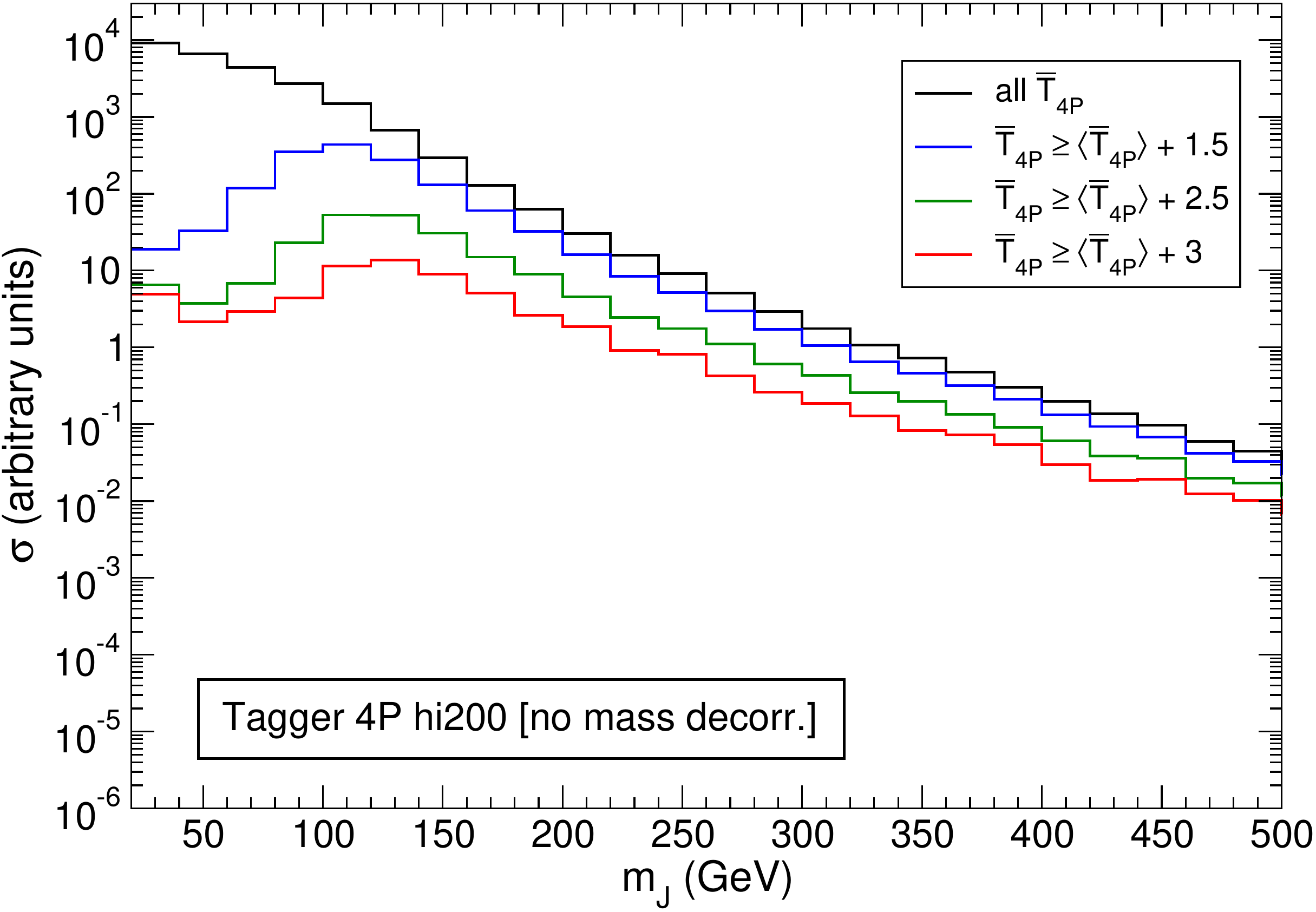} 
\end{tabular}
\caption{Jet mass distribution of the QCD background after increasingly tighter cuts on the $\bar T_\text{4P}$ taggers (without mass decorrelation) designed on two different jet mass intervals.}
\label{fig:mJT0}
\end{center}
\end{figure}

\begin{figure}[htb]
\begin{center}
\begin{tabular}{ccc}
\includegraphics[height=4.8cm,clip=]{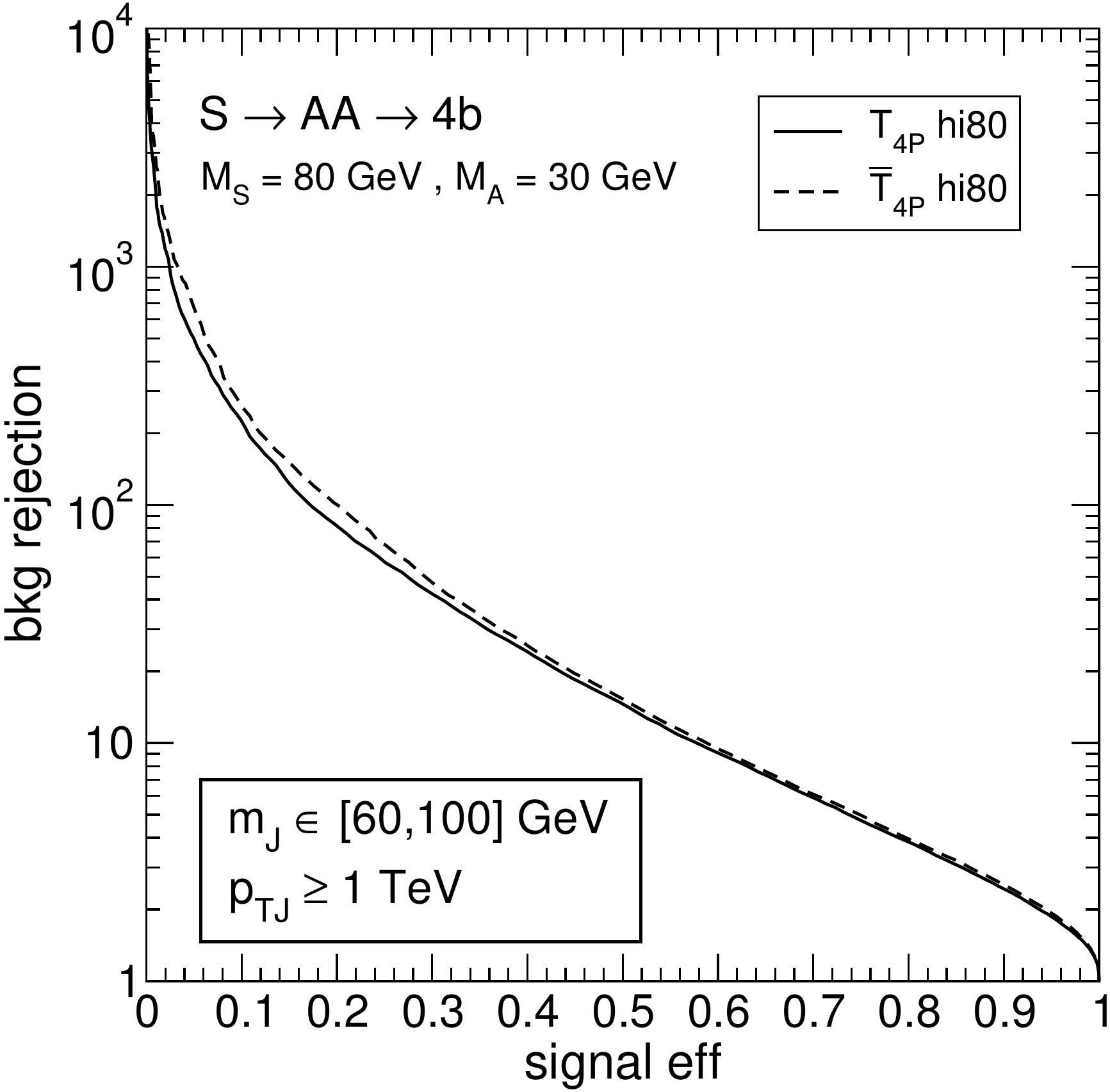} & 
\includegraphics[height=4.8cm,clip=]{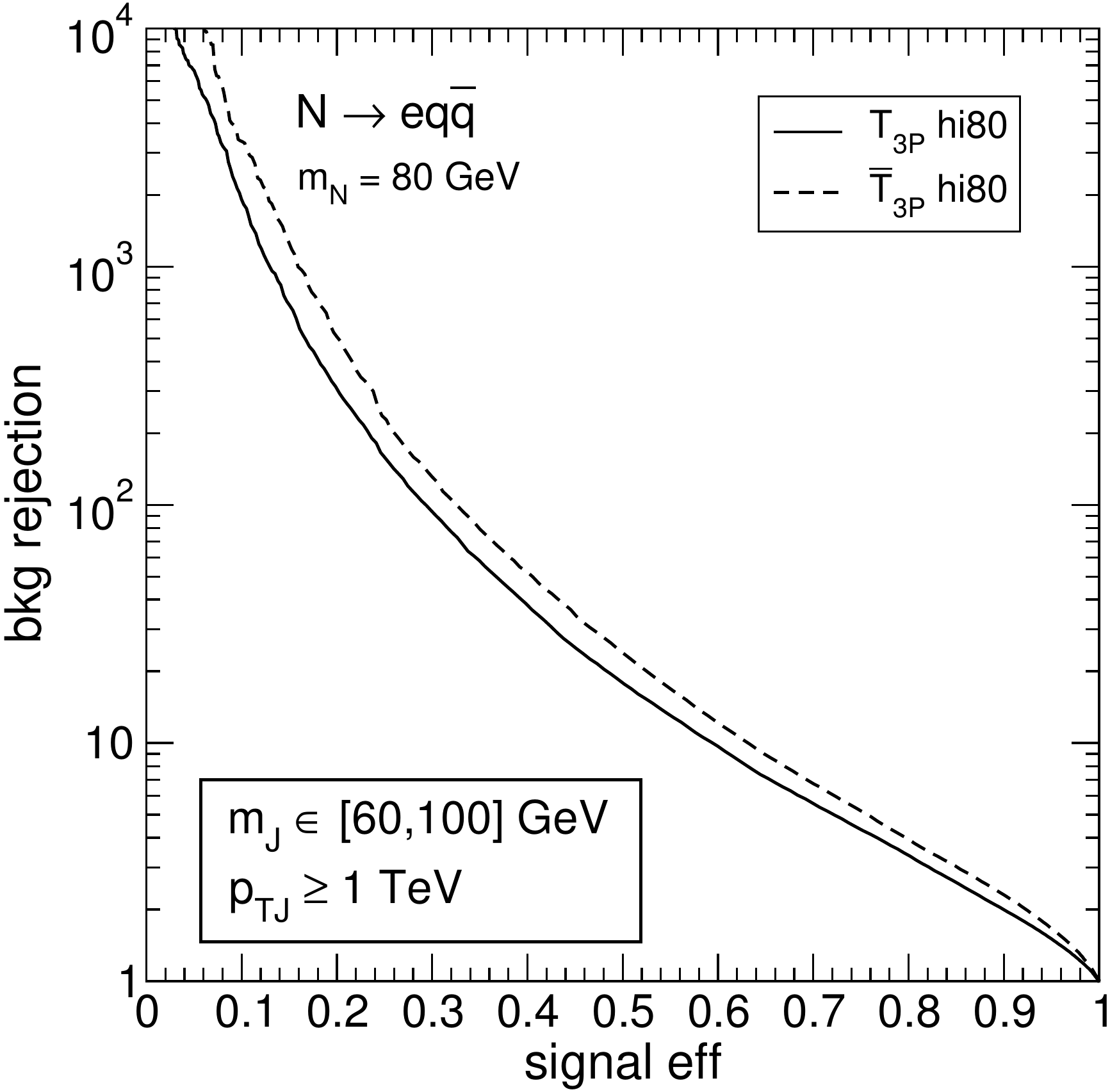} &
\includegraphics[height=4.8cm,clip=]{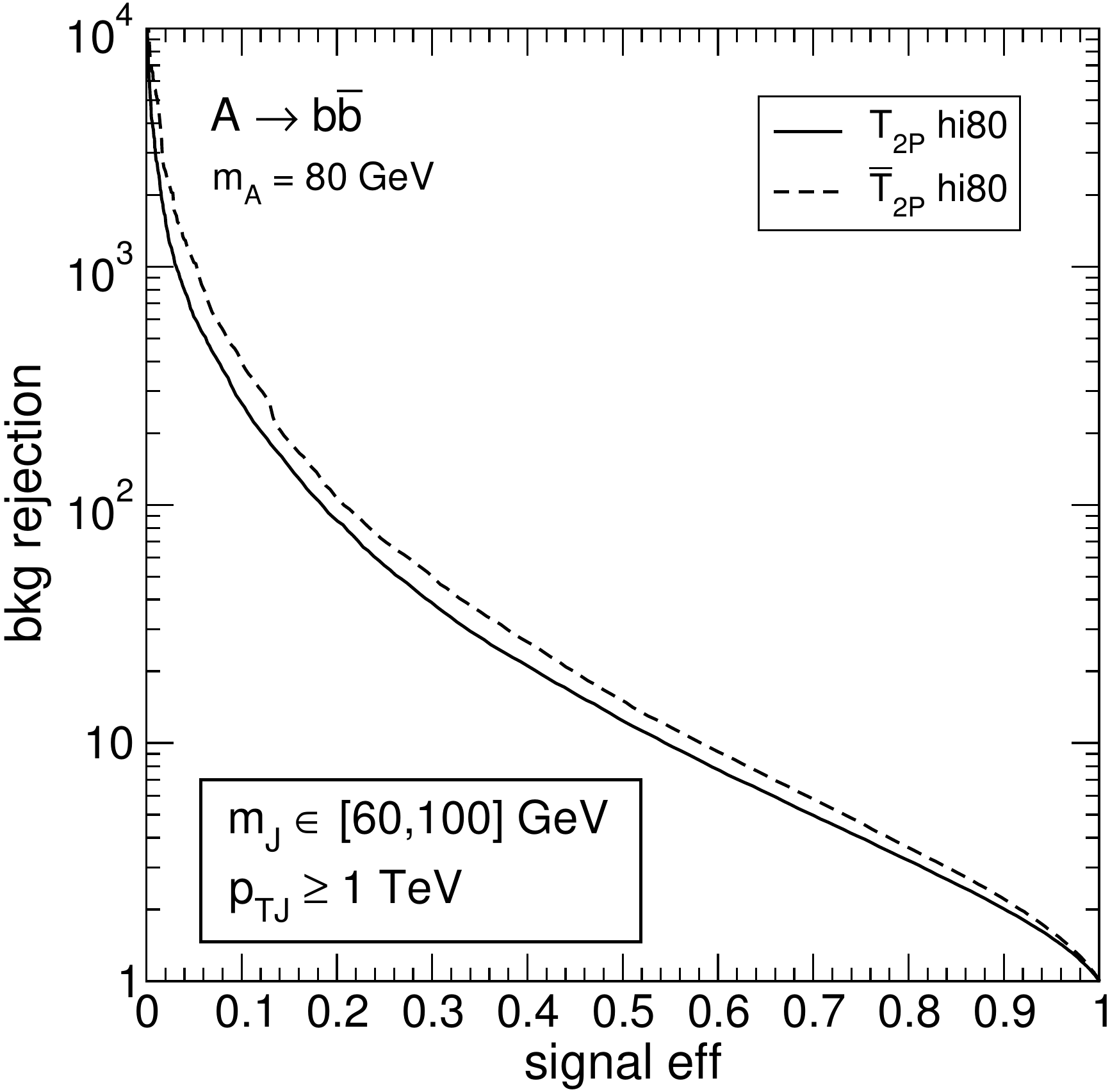} 
\end{tabular}
\caption{Performance of the taggers $T$ with mass decorrelation  versus the taggers $\bar T$ directly obtained from the LoRD.}
\label{fig:ROC-decorr}
\end{center}
\end{figure}

The application of the taggers without prior mass decorrelation produce a significant shaping of the jet mass spectrum for the QCD background. This is illustrated in Fig.~\ref{fig:mJT0} for two taggers $\bar T_\text{4P}$ designed in two different mass intervals. The peak-like structure produced is near 100 GeV in both cases, therefore the location of the bump is not related to the design mass interval, $[60,100]$ GeV for {\tt hi80} and $[170,230]$ GeV for {\tt hi200}. These examples highlight the benefit of the mass decorrelation even if it is not perfect, see Fig.~\ref{fig:mdecorr}.
A drawback of the decorrelation procedure is a slightly worse discrimination between signal and background, as it can be seen in Fig.~\ref{fig:ROC-decorr}, for four-pronged (left), three-pronged (middle) and two-pronged (right) taggers, designed in the same kinematical region {\tt hi80}.

\section{4P as generic taggers}
\label{sec:b}

\begin{figure}[htb]
\begin{center}
\begin{tabular}{ccc}
\includegraphics[height=5.2cm,clip=]{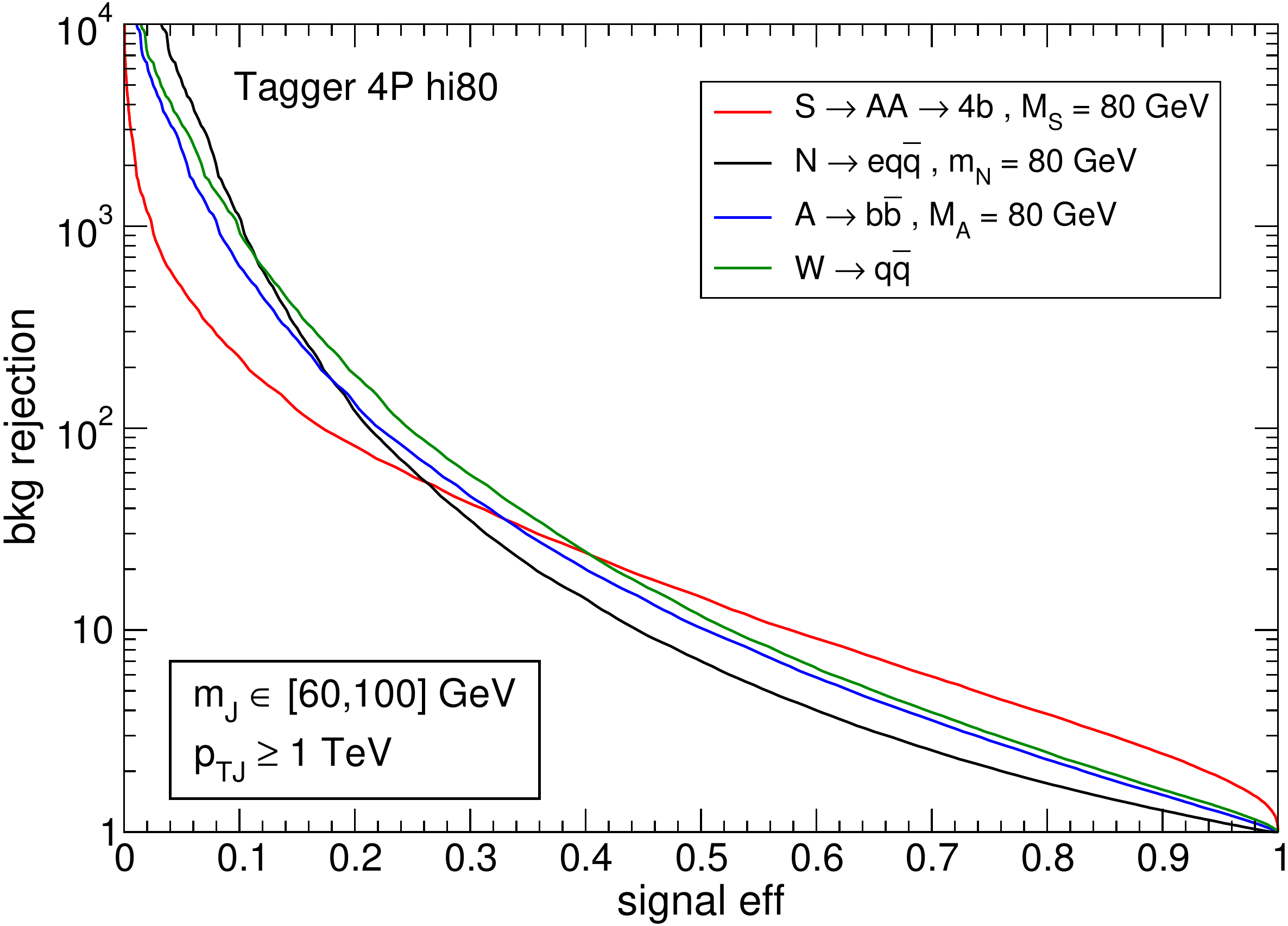} & \quad &
\includegraphics[height=5.2cm,clip=]{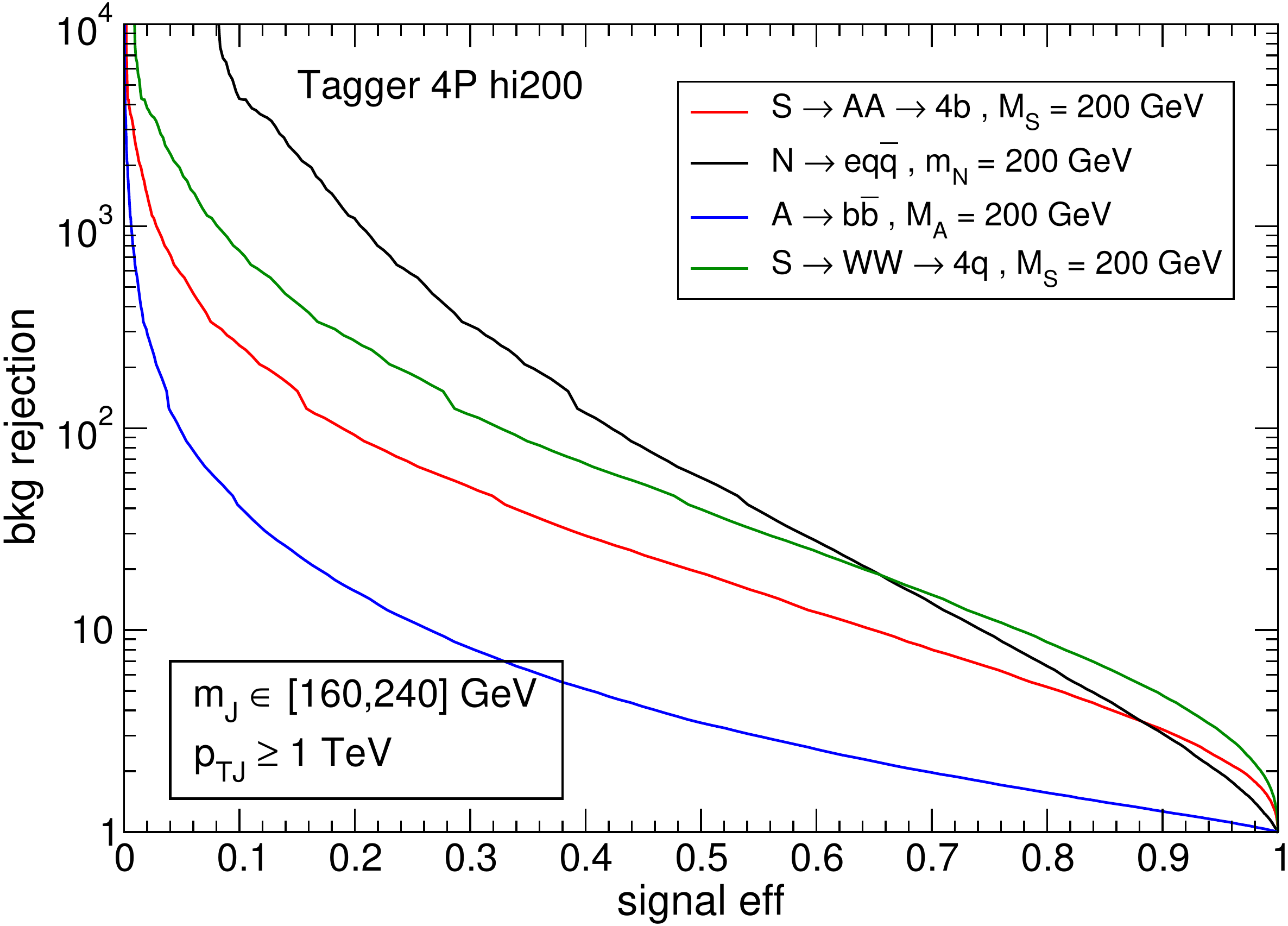}
\end{tabular}
\caption{Comparison of the performance of 4P {\tt hi80} (left) and {\tt hi200} (right) taggers  for several signals, from two-pronged to four-pronged.}
\label{fig:4Pasgen}
\end{center}
\end{figure}

Despite being specifically designed for four-pronged signals, the 4P {\tt hi80} and {\tt hi200} taggers work well for jets with less than four quarks. One example has been already shown in Fig.~\ref{fig:ROC1b}: jets containing two $b$ quarks and two photons. For completeness, we show here the performance for other signals : $N \to e q \bar q$, $A \to b \bar b$, and $W \to q \bar q$. The results are shown in Fig.~\ref{fig:4Pasgen}. We also include four-pronged signals as a reference. The masses of the particles originating the jet are the same as taken in Figs.~\ref{fig:ROC} and \ref{fig:ROC1b}. 

For low masses, the 4P {\tt hi80} tagger performs well for all signals --- actually, the one for which the discrimination is worse is the difficult four-pronged signal $S \to AA \to 4b$ for which it is specifically designed. For high masses, the 4P {\tt hi200} tagger is not adequate for $A \to b \bar b$ but it is quite good for $N \to e q \bar q$.

\end{document}